\documentclass[onecolumn,%
10pt,tightenlines,
notitlepage,%
nofootinbib,%
superscriptaddress,%
amsfonts,amsmath]{revtex4}

\usepackage{hyperref}
\usepackage{xcolor}

\newcommand{\del}{\partial}

\newcommand{\la}{\lambda}
\newcommand{\m}{\mu}
\newcommand{\n}{\nu}

\newcommand{\p}{\phi}
\newcommand{\vp}{\varphi}

\newcommand{\td}{\text{d}}
\newcommand{\ma}{\mathcal{A}}

\begin{document}

\title {Vainshtein screening for slowly rotating stars}

\author{T.~Anson}
\affiliation{Universit\'e Paris-Saclay, CNRS/IN2P3, IJCLab, 91405 Orsay, France}

\author{E.~Babichev}
\affiliation{Universit\'e Paris-Saclay, CNRS/IN2P3, IJCLab, 91405 Orsay, France}
\affiliation{Sorbonne Universit\'e, CNRS, UMR7095,
Institut d'Astrophysique de Paris,
${\mathcal{G}}{\mathbb{R}}\varepsilon{\mathbb{C}}{\mathcal{O}}$,\\
98bis boulevard Arago, F-75014 Paris, France}

\begin{abstract}
We study the Vainshtein mechanism in the context of slowly rotating stars in scalar-tensor theories. 
While the Vainshtein screening is well established for spherically symmetric spacetimes, we examine its validity in the axisymmetric case for slowly rotating sources.
We show that the deviations from the general relativity solution are small in the weak-field approximation outside the star: the solution for the frame-dragging function is the same as in general relativity at leading order. 
Moreover, in most cases the corrections are suppressed by powers of the Vainshtein radius provided that the screening operates in spherical symmetry. 
Outside the Vainshtein radius, the frame dragging function receives corrections that are not suppressed by the Vainshtein radius, but which are still subleading.  
This suggests that the Vainshtein mechanism in general can be extended to slowly rotating stars and that it works analogously to the static case inside the Vainshtein radius.
We also study relativistic stars and show that for some theories the frame-dragging function in vacuum does not receive corrections at all,
meaning that the screening is perfect outside the star.
\end{abstract}

\maketitle

\section{Introduction}

A way to test the validity of general relativity (GR) is to put constraints on theories that deviate from it. 
One of the approaches to modify GR is to add extra fields mediating the gravitational force, and the simplest extensions are scalar-tensor theories with one additional scalar field. 
However, GR passes all local experimental tests, therefore it is necessary to have a mechanism that screens the effect of the scalar field (fifth force) close to the gravitational source, i.e. in the Solar System.
Such a mechanism, analogous to the Vainshtein mechanism in the decoupling limit of massive gravity~\cite{Vainshtein:1972sx,Babichev:2009us,Babichev:2009jt,Babichev:2010jd} (see also~\cite{Alberte:2010it,Koyama:2011xz,Koyama:2011yg,Chkareuli:2011te,Volkov:2012wp,Babichev:2013pfa} and a review~\cite{Babichev:2013usa}), allows one to recover GR inside a so-called Vainshtein radius, while deviations from GR may be observed at large radii~\cite{Babichev:2009ee}. 
The Vainshtein mechanism has been extensively studied in scalar tensor theories for spherically symmetric spacetimes, 
in particular, in Horndeski~\cite{Babichev:2009ee,Kimura:2011dc,DeFelice:2011th,Babichev:2012re,Koyama:2013paa,Kase:2013uja,Chagoya:2014fza,Babichev:2016kdt} and beyond Horndeski~\cite{Kobayashi:2014ida,Kase:2015zva,Koyama:2015oma,Saito:2015fza,Babichev:2016jom,Sakstein:2016oel} theories, and in degenerate higher order scalar-tensor (DHOST) theories~\cite{Dima:2017pwp,Langlois:2017dyl,Crisostomi:2017lbg,Hirano:2019scf,Crisostomi:2019yfo}.

However, realistic astrophysical objects typically rotate, and one may naturally ask whether rotation affects the validity of the Vainshtein mechanism. 
Indeed, it has been found that the chameleon screening mechanism is shape dependent~\cite{Burrage:2017shh,Burrage:2014daa}, i.e. the fifth force does depend on the deviation from spherical symmetry. 
In the case of the Vainshtein mechanism, the recovery of GR for nonspherical configurations in particular models
has been previously considered in~\cite{Hiramatsu:2012xj,Chagoya:2014fza,Cisterna:2016vdx,Sakstein:2016oel}.

The aim of this work is to make a systematic study of the Vainshtein screening in scalar-tensor theories for slowly rotating bodies.
We consider generic quadratic DHOST Ia theories, meaning that the Lagrangian contains at most terms quadratic in the second derivatives of the scalar field, and can be mapped to the Horndeski theories via a general disformal transformation \cite{Achour:2016rkg}. 
To study the effects of slow rotation, we follow the Hartle-Thorne formalism developed for GR and include the scalar field in the discussion. 
We consider both time-dependent and static scalar fields, and derive the general equation for the frame-dragging function in a compact form. 
For a particular subclass of the DHOST theories with shift symmetry, we are able to establish the full recovery of GR in vacuum for slow rotation.
The rest of our results are found in the weak-field approximation, i.e. the metric is assumed to be almost flat, which allows us to make an expansion in small deviations from Minkowski spacetime. We study various cases of scalar-tensor theories and coupling to matter. 

The key feature of the Vainshtein mechanism can be most easily demonstrated for nonrelativistic spherically symmetric static configurations outside the source. 
The GR solution for the the metric is recovered inside the Vainshtein radius $r_V$ up to small corrections. 
More precisely, when the metric is written in the form
\begin{equation}
\text{d}s^2 = -e^{\nu(r)} \text{d}t^2 + e^{\lambda(r)} \text{d}r^2 + r^2 \text{d}\theta^2 + r^2\sin^2\theta\,\text{d}\varphi^2\; ,
\end{equation}
the GR vacuum solutions for the metric functions are recovered for distances smaller than the Vainshtein radius,
\begin{equation}
\label{metric_suppression}
\nu = -\frac{r_S}{r}\left[1+ \mathcal{O}\left(\frac{r}{r_V}\right)^n\right],\quad 
\lambda = \frac{r_S}{r}\left[1+ \mathcal{O}\left(\frac{r}{r_V}\right)^n\right],
\end{equation}
where $r_S$ is the Schwarzschild radius, $r_V$ is the Vainshtein radius and $n$ is a parameter which depends on the theory at hand. 
The deviations from flat spacetime are proportional to $r_S/r$ (as in GR) and the corrections to these deviations are suppressed due to the Vainshtein mechanism.  
Note that when the Vainshtein mechanism does not operate, normally the corrections to GR
are of the same order as the solution itself, i.e. instead of the Vainshtein suppressed terms one finds corrections of order $r_S/r$ in the above expressions.

A similar picture can be depicted for slow rotation. While $\nu$ and $\lambda$ are not modified, 
an extra metric function $\omega(r)$, which we dub the frame-dragging function, is added to take into account the effects of rotation. 
As we find later in the paper, 
the equation for $\omega$ outside the source in the weak-field approximation can be presented in the form 
\begin{equation*}
\frac{\td^2 \omega}{\td r^2} +\frac{4}{r}\left(1 +  \mathcal{S}\right) \frac{\td \omega}{\td r}=0\; ,
\end{equation*} 
where the term $\mathcal{S}$ shows modifications with respect to the GR, for which $\mathcal{S}=0$. 
In the weak-field approximation, we will see below that when the Vainshtein mechanism operates in spherical symmetry, then generically 
$\mathcal{S} =  \mathcal{O}\left(\frac{r}{r_V}\right)^l,$ 
where $l$ depends on the theory\footnote{
For some theories it can also happen that $\mathcal{S}$ is exactly zero even for relativistic stars, as we show below.}. 
Then, the solution for $\omega$ acquires leading corrections of the same order as $\mathcal{S}$ , which is analogous to the screening for nonrotating sources (see Eq.~(\ref{metric_suppression})).
On the other hand, we have $\mathcal{S}= \mathcal{O}\left(r_S/r\right)$ in the non-Vainshtein regime, 
i.e. $\omega$ receives corrections of order $r_S/r$, while the leading term is not modified.
This is different from what happens for the metric functions $\{\lambda,\nu\}$, whose leading term is modified when the Vainshtein screening no longer operates.

Inside matter, the situation is more complicated, and the Vainshtein mechanism for rotating sources can be broken. 
We will say that the Vainshtein screening for the frame-dragging function operates if the leading term of the solution for $\omega$ is the same as in GR.
In situations where the screening works, we examine the leading corrections to the frame dragging function in the weak-field limit. 
We will see that that the screening for $\omega$ is usually more effective (meaning that corrections to the GR solution are suppressed by powers of $r_V$) when the Vainshtein mechanism operates in the nonrotating case. 
However, we show some examples for which this is not true.

The plan of the paper is the following. In section~\ref{sec:setup} we give the action for DHOST Ia theories, describe the Hartle-Thorne formalism and derive equations of motion for slowly rotating sources. 
In section~\ref{sec:relativistic} we apply the formalism to shift-symmetric DHOST Ia Lagrangians and find, in some cases, the full recovery of GR in vacuum, i.e. the equation for the frame-dragging function is exactly the same as in GR. 
In the following section \ref{sec:solutions},  we assume the weak-field approximation in addition to slow rotation. 
We examine the equation for the frame-dragging function, and find leading and subleading terms of the solution in the general form.
In section~\ref{sec:weakfield1}, we systematically study the effects of slow rotation in general scalar-tensor theories for a time-dependent scalar field. We consider several subcases, depending on the structure of the equations in the nonrotating limit, and study subleading corrections to the frame-dragging function. 
Section~\ref{sec:weakfield2} is devoted to the time-independent scalar field, where the nonzero fifth force is due to a nonminimal coupling of the scalar field to the curvature.
Finally, we conclude in section~\ref{sec:conclusion}.

\section{Action and equations of motion for slow rotation}
\label{sec:setup}

We will consider quadratic DHOST theories, meaning that the Lagrangian contains terms at most quadratic in second derivatives of the scala field. The action is given by~\cite{Crisostomi:2016czh,Langlois:2015cwa}
\begin{equation}
\label{actionDHOST}
S = M_P^2\int\td^4 x \sqrt{-g}\left( f(\phi,X)R+K(\phi,X)-G_3(\phi,X)\square\phi + \sum_{i=1}^{5}A_i(\phi,X)\mathcal{L}_i\right)+ S_\text{m}\left[g_{\mu\nu},\psi_\text{m}\right]\; ,
\end{equation}
 where $M_P=(8\pi G)^{-1/2}$ is the reduced Planck mass, $\phi$ is a dimensionless scalar field and $X = -\frac12 (\del\phi)^2$. 
Defining $\phi_\mu = \nabla_\mu\phi$ and $\phi_{\m\n}=\nabla_\mu\nabla_\nu\phi
$, the $\mathcal{L}_i$ are given by
\begin{equation}
\label{L}
\mathcal{L}_1=\p_{\m\n}\p^{\m\n}, \quad 
\mathcal{L}_2=(\square\phi)^2, \quad 
\mathcal{L}_3=\phi_{\mu\n}\p^\m\p^\n \square\phi, \quad
\mathcal{L}_4= \phi_\mu\phi^\nu\phi^{\mu\alpha}\phi_{\nu\alpha}, \quad
\mathcal{L}_5= \left(\phi_{\mu\n}\p^\m\p^\n\right)^2\; .
\end{equation}
The most interesting case of the above action is the DHOST Class Ia~\cite{Langlois:2015cwa}. 
It is obtained by imposing three constraints on the functions $A_i$. Assuming
$f+2XA_1\neq 0$, one can express $A_2, A_4, A_5$ in terms of $f, A_1, A_3$ as follows:
\begin{equation*}
\begin{split}
A_2 &= -A_1\; ,\\
A_4 &= \frac{8 X A_1^3 + A_1^2(3f + 16 X f_X) - X^2 f A_3^2 + A_3 A_1 (8 X^2 f_X - 6 X f)+ 2f_XA_1(3f + 4X f_X)+2 f A_3(Xf_X-f)+3 ff_X^2}{2(f+2X A_1)^2},\\
A_5 &= \frac{(f_X + A_1 + X A_3)(A_1^2-3X A_1 A_3 + f_XA_1 - 2fA^3)}{2(f+2X A_1)^2}\; ,
\end{split}
\end{equation*}
where the subscript $X$ should be understood as the derivative with respect to $X$, i.e. $f_X\equiv \del f/\del X$, etc.
Notice that the above expressions differ from those in~\cite{Langlois:2015cwa} because of our definition of $X$.
It is known that in spherical symmetry, theories belonging to this class exhibit the Vainshtein screening~\cite{Dima:2017pwp}, meaning that GR is recovered inside a radius $r_V$ called the Vainshtein radius, and deviations from GR may be observed at large radii. For some theories beyond Horndeski, the screening is broken inside matter~\cite{Kobayashi:2014ida} when the scalar field depends on time, and sometimes even outside the matter source \cite{Crisostomi:2019yfo,Hirano:2019scf}. 
In this work, we will extend these studies by deviating from spherical symmetry and examining how the Vainshtein screening is affected.

We consider a slowly rotating source of radius $R$ modeled by a perfect fluid. We will follow the Hartle-Thorne formalism \cite{Hartle:1967he} developed for general relativity, and assume a uniform rotation of the fluid at angular velocity $\Omega$. 
We take the same ansatz for the metric tensor as in GR,
\begin{equation}
\label{HTmetric}
\text{d}s^2 = -e^{\nu(t,r)} \text{d}t^2 + e^{\lambda(t,r)} \text{d}r^2 + r^2 \text{d}\theta^2 + r^2\sin^2\theta\left[\text{d}\varphi - \varepsilon\omega(t,r) \text{d}t\right]^2\; ,
\end{equation}
 where the frame-dragging function $\omega$ is the angular velocity acquired by an observer falling freely from infinity, due to the dragging of inertial frames. The bookkeeping parameter $\varepsilon$ accounts for the slow rotation of the source, and we will keep only terms up to first order in $\varepsilon$ in the following. 
The relation between the functions $\{\la,\nu\}$ used in this work and the Newtonian potentials $\{\Phi,\Psi\}$ often encountered in the literature can be found in Appendix~\ref{appendix:b}. For the scalar field we take the (generically) time-dependent ansatz~\cite{Babichev:2010kj,Babichev:2012re,Babichev:2013cya}
 \begin{equation}
 \label{phians}
\phi = q t + \phi(r)\;.
\end{equation}
 The metric functions can \textit{a priori} depend on time if the constant $q\neq0$. Indeed, the Lagrangian functions generically depend on $\phi(t)$. The solutions for the metric potentials depend on these functions, meaning that they also depend on time. We assume the energy-momentum tensor of the perfect fluid
 \begin{equation}
 \label{energy_momentum}
 T^{\m\n} = \left(\rho+ P\right)u^\mu u^\nu + P g^{\m\n}\; ,
 \end{equation}
 where $u^\mu$ is the $4$-velocity of the fluid,  given at first order in $\varepsilon$ by
 \begin{equation}
 \label{4vel}
 u^\mu = \left(e^{-\nu/2},0,0,\varepsilon\Omega e^{-\nu/2}\right)\; .
 \end{equation}
We will calculate the equations of motion up to  order 1 in $\varepsilon$. We will be interested in the differential equation for the function $\omega$, obtained from the $t\varphi$ component of the metric equations:
\begin{equation}
\label{tphi}
{\mathcal{E}^t}_\varphi=\frac{1}{2M_P^2} {T^t}_\varphi\; ,
\end{equation}
where $\mathcal{E}_{\m\n} =\frac{1}{\sqrt{-g}} \frac{\delta \left(\sqrt{-g}\mathcal{L}\right)}{\delta g^{\m\n}}$ and $T_{\mu\nu}=-\frac{2}{\sqrt{-g}}\frac{\delta \left(\sqrt{-g}\mathcal{L}_\text{m}\right)}{\delta g^{\m\n}}$. 
At the same time, for the other nontrivial equations, $(tt)$, $(rr)$ and $(tr)$ components as well as for the scalar field equation, it is enough to keep zero order in $\varepsilon$,
i.e. to consider these equations of motion without rotation,
\begin{eqnarray}
\mathcal{E}^{\text{(st)}}_{tt}&=&\frac{1}{2M_P^2} T^{\text{(st)}}_{tt}\; ,\label{Emetric00}\\
\mathcal{E}^{\text{(st)}}_{rr}&=&\frac{1}{2M_P^2} T^{\text{(st)}}_{rr}\; ,\label{Emetric11} \\
\mathcal{E}^{\text{(st)}}_{tr}&=&0\; , \label{Emetric01}\\
\mathcal{E}^{\text{(st)}}_{\phi} &=& 0\; ,\label{Ephi}
\end{eqnarray}
where the superscript (st) implies that one should set $\varepsilon =0$ in the equations of motion.
Note that not all of the equations~(\ref{Emetric00})--(\ref{Ephi}) are independent, because of the following relation due to the diffeomorphism invariance of the action:
\begin{equation}
\nabla^\nu\mathcal{E}_{\mu\nu}  = -\frac{1}{2}\nabla_\mu\phi\;  \mathcal{E}_\phi\; .
\end{equation}
With the choice (\ref{energy_momentum}) for $T_{\mu\nu}$, Eq.~(\ref{tphi}) can be written as
\begin{equation}
\label{tpeq}
\omega'' + K_1 \omega' + \frac{K_2}{M_P^2} \left(\rho+P\right)\left(\omega-\Omega\right)=0\; ,
\end{equation}
where the functions $K_1$ and $K_2$ depend on the specific theory considered and on the solution in the nonrotating limit,
\begin{eqnarray}
\label{K1full}
K_1&=&\frac4r -\frac{\lambda'+\nu'}2 + \frac{\td}{\td r}\ln\left(f+2XA_1\right)\; ,\\
\label{K2full}
K_2&=&-\frac{e^{\lambda}}{f+2XA_1}\; ,
\end{eqnarray} 
and $'$ denotes a derivative with respect to the radial coordinate.
Thus the system of equations to solve is given by (\ref{Emetric00})--(\ref{Ephi}) and (\ref{tpeq}) with (\ref{K1full}) and (\ref{K2full}), 
where all the functions depend on $\phi$ given by (\ref{phians}) and $X$ evaluated in the spherically symmetric limit,
\begin{equation}
\label{X}
X= \frac12 \left(e^{-\nu} q^2- e^{-\lambda} \phi'^2\right).
\end{equation}
Using (\ref{X}),  Eq.~(\ref{K1full}) can be written in an expanded form, which will be useful in the following, as
\begin{equation}
\label{K1fullbis}
K_1=\frac4r -\frac{\lambda'+\nu'}2 + 
\frac{\left(2 X A_{1X} + 2 A_1 + f_X \right)X'
+ \phi'\left( 2 XA_{1\phi} + f_\phi \right)}{2XA_1+f}\; .
\end{equation}
where 
\begin{equation}
\label{Xp}
X'=\frac12\left[e^{-\lambda}\phi'\left(\lambda'\phi'-2\phi''\right)-q^2e^{-\nu}\nu'\right]\; .
\end{equation}
Eq.~(\ref{tpeq}) with the coefficients given by~(\ref{K1full}) and~(\ref{K2full}) is the main equation we will focus on throughout the paper. 

Note that the GR case is easily obtained from the above equations. Indeed, we set $\mathcal{L}=R/2$, corresponding to $G_3=K=A_i=0$ and $f=1/2$. 
Using~(\ref{K1full}) and (\ref{K2full}) in~(\ref{tpeq}) one obtains
\begin{equation}
\label{omegaGR}
\omega'' + \left(\frac{4}{r}-\frac{\la'+\nu'}{2}\right)\omega' - \frac{2}{M_P^2}e^\la\left(\rho+P\right)\left(\omega-\Omega\right)=0\; .
\end{equation}
which coincides with the GR equation for $\omega$ \cite{Hartle:1967he}. 
In vacuum we impose $\rho=P=0$, which implies $\lambda' = -\nu'$ in GR, so that Eq.~(\ref{omegaGR}) becomes
\begin{equation}
\label{omegaGRvac}
\omega''+\frac{4}{r}\omega'=0\; .
\end{equation}
The solution to this equation is
\begin{equation}
\label{omegasolGR}
\omega =\frac{2JG}{r^3}\; ,
\end{equation}
where $J$ is the total angular momentum of the star \cite{Papapetrou:1948jw,Hartle:1967he}, and we have set $\lim_{r\to\infty}\omega=0$. The angular momentum can be expressed in terms of the moment of inertia $I$ of the star as $J=\Omega I$.
In the following, we will examine the solutions for $\omega$ in DHOST Ia theories and compare them to the GR expression Eq.~(\ref{omegasolGR}).


\section{Slow rotation of relativistic sources in shift-symmetric theories}
\label{sec:relativistic}
In this section, we study the slow rotation of relativistic stars for shift-symmetric theories that are invariant under $\phi\to -\phi$, meaning we set $G_3=0$ throughout this section.  
We also assume slow rotation, but otherwise the equations are fully nonlinear in the metric functions $\lambda$ and $\nu$, i.e. we do not assume the weak-field approximation in this section.

\subsection{Horndeski theories}
We first consider Horndeski theories with general functions $f(X)$ and $K(X)$. The Lagrangian density reads:
\begin{equation}
\label{Horndeski}
\mathcal{L} =K(X) + f(X) R +f_X\left[(\square\phi)^2-\phi_{\mu\nu}\phi^{\mu\nu}\right]\; .
\end{equation}
The authors of~\cite{Cisterna:2016vdx} studied slowly rotating neutron stars in the case when $f(X)$ and $K(X)$ are linear functions of $X$. 
They showed that the equation for $\omega$ in vacuum reduced to the GR expression, meaning that we have $K_1=4/r$ and the term proportional to $K_2$ in Eq.~(\ref{tpeq}) is absent. 
We extend this result to a more general class of theories.
We assume $f_{XX}\neq 0$, while the case $f_{XX}=0$ was treated in \cite{Cisterna:2016vdx}. 
With this assumption, the scalar field can be obtained in terms of $\{\la,\nu,\nu'\}$ from the equation $\mathcal{E}_{tr}=0$:
\begin{equation*}
\p'^2 = \frac{e^\la\left[2f_X\left(1+r\n' -e^\la\right) + r\left(2q^2f_{XX}\n'e^{-\n}-rK_Xe^{-\la}\right)\right]}{2f_{XX}\left(1+r\n'\right)}\; .
\end{equation*} 
One then substitutes this expression into the $(rr)$ component of the metric equations to obtain $\la$ in terms of $\n'$:
\begin{equation*}
e^\la = \frac{2\left(1+r\n'\right)\left(f_X^2+ff_{XX}\right)}{2f_X^2+r^2f_XK_X + f_{XX}\left(2f + r^2K + r^2 P/M_P^2\right)}\; .
\end{equation*}
Using the $(tt)$ equation one can then obtain $\la'$ in terms of $\{\la,\nu,\p',\p'',\rho\}$. 
After substituting this expression in Eq.~(\ref{K1fullbis}), the second derivatives of $\phi$ disappear and we are left with a coefficient $K_1$ which depends only on $\{\la,\nu,\nu',\p'^2\}$. 
Upon substituting the expressions for $\phi'^2$ and $\la$ the final expression for $K_1$ depends only $\{\rho,P,\nu,\nu'\}$. 
Finally, the coefficients read:
\begin{equation*}
\begin{split}
	K_1 &= \frac{4}{r}-\frac{re^\nu(1+r\n')^2(f_X^2+ff_{XX})(\rho+P)}{2M_P^2e^\n[2f+r^2(P/M_P^2+K)](f_X^2+ff_{XX})(1+r\n')-2q^2f_X[2f_X^2+r^2f_XK_X+[2f+r^2(K+P)]f_{XX}]}\; ,\\
	K_2 &=- \frac{e^\la}{f-2 X f_X}\; .
	\end{split}
\end{equation*}
One can see that the GR case is recovered in vacuum, where we simply have $K_1=4/r$. This shows that the result of \cite{Cisterna:2016vdx} can be extended to general functions $f$ and $K$ in Horndeski theories. 

It is also worth pointing out a mistake in formulas~(44) and~(53) of Ref.~\cite{Cisterna:2016vdx}. 
In their notations [obtained from ours by $\omega\to\Omega_*-\omega$, $e^\n\to b$, $q\to Q$ and $K_2\to -K_2 (\rho+P)$], these formulas should read
\begin{equation*}
\begin{split}
	u_\vp & = \varepsilon\frac{r^2\sin^2\theta\omega}{\sqrt{b}}\; ,\\
	K_2 &= \frac{4(b+rb')^2(P+\rho)}{b[(P r^2+4\kappa)(b+rb')-\eta Q^2]}\; .
\end{split}
\end{equation*}
With the above expression for $u_\vp$, one recovers the correct expression for the 4-velocity vector~\cite{Hartle:1967he}:
\begin{equation*}
	u^\mu = \left(u^0,0,0,\epsilon \Omega u^0\right)\; ,
\end{equation*}
unlike the case of Ref.~\cite{Cisterna:2016vdx}.

\subsection{Quadratic GLPV theories}
The above result---namely, that the equation for $\omega$ reduces to the one of GR in vacuum---shown for Horndeski theory with arbitrary $f(X)$ and $K(X)$ can be extended to quadratic Gleyzes-Langlois-Piazza-Vernizzi (GLPV) theories. We consider the following Lagrangian density~\cite{Gleyzes:2014dya}:
\begin{equation}
\label{BHtheory}
	\mathcal{L} =  K(X)+ f(X) R +f_X\left[(\square\phi)^2-\phi_{\mu\nu}\phi^{\mu\nu}\right] + \frac{A_3(X)}{2}\varepsilon^{\mu\nu\alpha\sigma}{\varepsilon^{\lambda\eta\kappa}}_\sigma\phi_{\mu\lambda}\phi_{\nu\eta}\phi_\alpha\phi_\kappa\; ,
\end{equation}
where $\varepsilon^{\mu\nu\alpha\sigma}$ is the totally antisymmetric Levi-Civita tensor, and we have set
\begin{equation}
	\label{GLPVdef}
	f = f(X), \quad
	A_1 = -A_2 = -f_X - X A_3(X),\quad
	A_4 = -A_3 (X), \quad
	K = K(X),\quad  
	A_5 =G_3 = 0\; .
\end{equation}
The inclusion of $A_3$ makes $\mathcal{E}_{tr}=0$ a quadratic equation in $\phi'^2$, in contrast to the Horndeski case, where the analogous equation is linear in $\phi'^2$. 
In order to obtain the desired result, we use the metric equations in a different order than in the previous case for Horndeski theory. 
First, we use $\mathcal{E}_{rr}$ to express $\phi'\phi''$ in terms of $\{\phi',\la,\nu,\nu'\}$. Then, we substitute this expression into $\mathcal{E}_{tt}$ to obtain $\la'$ in terms of $\{\phi',\la,\nu,\nu'\}$, which we inject into $\mathcal{E}_{tr}$. This yields a quadratic equation for $\phi'^2$, and the two solutions are expressed in terms of $\{\la,\nu,\nu'\}$. 
Using the expressions for $\{\phi'\phi'', \la', \phi'^2\}$, one can obtain that $K_1=4/r$ in vacuum, which means that the GR equation for $\omega$ is fully recovered in the case of~(\ref{BHtheory}) as well. 

\subsection{DHOST Ia with constant \texorpdfstring{$X$}{X}}

Assuming in addition constant $X$, i.e. $X_0=q^2/2$,  the previous result can be extended to shift-symmetric DHOST theories. 
Indeed, when $X=$const., $A_4$ and $A_5$, defined in~(\ref{L}) drop out of the field equations, because one can rewrite them as 
\begin{equation*}
	\mathcal{L}_4 = X_\mu X^\mu\; , \quad \mathcal{L}_5 = \left(X_\mu \phi^\mu\right)^2 \; .
\end{equation*}
Since the above expressions are quadratic in $X_\mu$, their variation will not give any contribution to the field equations when $X$ is constant.
Then it immediately follows from~(\ref{K1full}) that
\begin{equation*}
	K_1 = \frac{4}{r}-\frac{\la'+\nu'}{2}\; ,
\end{equation*}
since $f(X_0)+2X_0 A_1(X_0)$ is a constant.
 With the choice $X_0=q^2/2$, the scalar can be expressed in terms of $\{\la,\nu\}$ as
\begin{equation}
\label{phip}
\phi'^2 = q^2e^\la\left(e^{-\nu}-1\right)\; .
\end{equation}
Using~(\ref{phip}) in the $tt$, $tr$ and $rr$ components of the metric equations, one can show that
\begin{equation*}
\la'+\nu'\sim r\left(P+\rho\right)\; ,
\end{equation*}
so once again the GR equation for $\omega$, Eq.~(\ref{omegaGRvac}), is recovered in vacuum.
\section{Weak-field approximation: equation for the frame-dragging equation and its solutions}
\label{sec:solutions}
From now on we will employ the weak-field approximation~\cite{Babichev:2010jd}, assuming
that $\lambda$, $\nu$, $\phi$ and their derivatives are small, which one can check once the solutions are found:
\begin{equation}
\label{cond1}
\{r^n\frac{\td^n\la}{\td r^n},r^n\frac{\td^n\n}{\td r^n},r^n\frac{\td^n\p}{\td r^n}\}\ll 1 \; ,
\end{equation}
where $n$ is a positive integer. Additionally, we assume that
\begin{equation*}
\omega\ll \Omega\; ,
\end{equation*}
which is the appropriate approximation in the Newtonian regime \cite{Hartle:1967he}. Physically, the above conditions correspond to nonrelativistic sources, for which we also assume $P\ll \rho$. These assumptions considerably simplify Eq.~(\ref{tpeq}), since it becomes a first order equation for $\omega'$:
\begin{equation}
\label{omegaeq_weak}
\omega'' +\frac{4}{r}\left[1 + \frac{r\delta K_1}{4}\right]\omega' = \frac{K_2(r)\Omega}{M_P^2}\rho(r)\; ,
\end{equation}
where 
$$\delta K_1 \equiv K_1-\frac{4}r$$
marks the departure from the vacuum GR behavior ($\rho=0$). The integration of  Eq.~(\ref{omegaeq_weak}) with the conditions $\omega'(0)=0$ and $\lim\limits_{r\to\infty}\omega=0$ leads to
\begin{equation}
\label{omega_weak}
\omega(r) = \frac{\Omega}{M_P^2}\int_{\infty}^{r}\frac{\mathcal{I}_1(v)}{v^4}\left(\int_{0}^{v}\frac{K_2(u) \rho(u)}{\mathcal{I}_1(u)}u^4 \td u\right) \td v\; ,
\end{equation}
where we have defined the function
\begin{equation*}
\mathcal{I}_1(r) = e^{-\int\delta K_1\td r}\; .
\end{equation*} 
We see that the overall integration constant for $\mathcal{I}_1$ is not important, as it disappears in the final result for  $\omega$. 
Note that~(\ref{omega_weak}) is valid for $\delta K_1$ not necessarily small.
In order to make a comparison of a generic situation with GR, let us briefly go through the GR case, i.e. $f=1/2$ and  $G_3=K=A_i=0$. The linearization of Eq.(\ref{Emetric00}) and (\ref{Emetric11}) gives, respectively,
\begin{equation*}
\begin{split}
\la+r \la' &= \frac{1}{M_P^2}r^2 \rho\; ,\\
r\nu' - \la &=\frac{1}{M_P^2} r^2 P\; .
\end{split}
\end{equation*}
In the case of nonrelativistic matter, $P\ll\rho$, we obtain from~(\ref{omegaGR}):
\begin{equation}
\label{omegaweakGR}
\omega'' + \frac{4}{r}\left(1-\frac{G M'}{4}\right)\omega'= -\frac{4GM' \Omega }{r^2}\left(1+\frac{2GM}{r}\right)\; ,
\end{equation}
where $M = 4\pi\int_{0}^{r}\rho u^2\td u$.
This is the weak-field equivalent of the relativistic GR equation found in~\cite{Hartle:1967he}.
As one can see by comparing~(\ref{omegaeq_weak}) and (\ref{omegaweakGR}), outside the source $\delta K_1$ measures the departure from GR, while inside the source it takes into account both GR and non-GR corrections due to the presence of matter.

\subsection{Leading term}
Let us now calculate the leading term in ($\ref{omega_weak}$), assuming that $\int \delta K_1 \td r$ is small and  $K_2$ is almost constant. We can then write
\begin{equation}
\label{epsilon_coeffs}
\begin{split}
\mathcal{I}_1 &= 1 + \varepsilon \delta\mathcal{I}_1\; ,\\
K_2 &= \kappa_2\left(1+ \varepsilon\delta K_2\right)\; ,
\end{split}
\end{equation}
where $\kappa_2$ is a constant, $\{\delta \mathcal{I}_1,\delta K_2\}\ll 1$, and $\varepsilon$ is a bookkeeping parameter used to keep track of small terms.
\paragraph{Outside the source:}
In the exterior region, $r>R$, Eq.~(\ref{omega_weak}) simplifies to
\begin{equation}
\omega(r) = \frac{2 G \tilde{J}}{r^3} + \mathcal{O}\left(\varepsilon\right)\; ,
\end{equation}
where we have defined an effective angular momentum 
\begin{equation}
\label{Jtilde}
\tilde{J}=- \frac{4\pi \Omega}{3}\int_{0}^{R}\frac{K_2(u) \rho(u)}{\mathcal{I}_1(u)}u^4 \td u \; .
\end{equation}
This coefficient can \textit{a priori} be different from the GR value. However, if the density profile of the star is unknown, $\mathcal{I}_1$ and $K_2$ can be reabsorbed in the definition of $\rho$.
Therefore, unless the density profile $\rho(r)$ is known, any physical effect related to frame-dragging outside the star is the same as in GR at leading order. Thus, one can say that the Vainshtein screening can be extended outside the star to the case of slowly rotating bodies in the weak-field approximation.
\paragraph{Inside the source:}
Inside the source, we have from~(\ref{omega_weak})
\begin{equation}
\label{omega_in_leading}
\omega-\omega(0)=\frac{\kappa_2 \Omega}{M_P^2}\int_{0}^{r}\frac{1}{v^4}\left(\int_{0}^{v}\rho(u)u^4 \td u\right)\td v +\mathcal{O}\left(\varepsilon\right)\; .
\end{equation}
The constant $\omega(0)$ is not free and it should be fixed by continuity at the surface of the star. 
One can see that for $\kappa_2\neq -2$, the solution for $\omega$ differs from its GR counterpart 
at leading order inside the star\footnote{Note that nonrotating solutions in some theories require a renormalization of $M_P$. 
In this case one should write~(\ref{omega_in_leading}) in terms of the renormalized Planck mass and take into account this extra factor in the definition of $\kappa_2$.}. In this case the Vainshtein mechanism is broken for rotating solutions inside the star.
On the other hand, the Vainshtein screening operates for theories in which $\kappa_2 = -2$ (for instance $A_1=0$ and $f=1/2$).
As an illustration, let us consider a constant density star with $\rho=\rho_0$ for $r<R$. From~(\ref{omega_in_leading}) we have
\begin{equation*}
\omega-\omega(0)=\frac{\kappa_2\rho_0\Omega r^2}{10 M_P^2} + \mathcal{O}\left(\varepsilon\right)\; .
\end{equation*} 
In order for $\tilde{J}$ to be positive at leading order in Eq.(\ref{Jtilde}), one must have $\kappa_2<0$. This implies that $\omega(r)$ is everywhere decreasing (as in GR) and that it is maximal  at $r=0$.

\subsection{Subleading terms}
In this subsection, we examine the subleading terms in the solution to Eq.~(\ref{omegaeq_weak}), when the corrections to the coefficients  $K_1$ and $K_2$ are power laws. The coefficient $K_2$ is only relevant inside the star where $\rho\neq 0$. 
On the other hand we will be interested in the corrections to $K_1$ for all $r$. 
As we will see in the following, one can in general identify three regions of radii, and in each of those the correction $\delta K_1$ has a particular power-law behavior. These regions are $r< R$, $R\leq r\ll r_V$ and $r\gg r_V$, where $r_V$ is the Vainshtein radius of the considered theory. Therefore, we can write approximately
\begin{align*}
\frac{r\delta K_1}{4}&=  a_1\left(\frac{r}{r_1}\right)^{s_1}H_{r\leq R} + a_2\left(\frac{r}{r_2}\right)^{s_2} H_{R< r\leq r_V} + a_3\left(\frac{r}{r_3}\right)^{s_3} H_{ r> r_V} \; , \\
\delta K_2 &=  a_0\left(\frac{r}{r_0}\right)^{s_0}\; ,
\end{align*}
where $H$ is the Heaviside function, $a_i$ are constants, and we assume that $(r/r_i)^{s_i}\ll 1$. The scaling exponents $s_i$ depend on the theory at hand and should satisfy certain constraints in order for the integral~(\ref{omega_weak}) to be finite and for $\omega$ to have the correct boundary conditions. Therefore we set $s_0+1> 0$, $s_1+1>0$, $s_2\neq0$ and $s_3<0$. We also assume $s_2\neq 3$, since we did not find an example of a theory with such a behavior, although it is not difficult to consider the case $s_2= 3$ separately.
It is worth noting that in the case of a time-dependent scalar field, Section~\ref{sec:weakfield1}, our analysis 
allows us to calculate the coefficients $K_1$ and $K_2$ up to $r\sim 1/q$.
In this case, instead of imposing the boundary condition at $r=+\infty$, we set the boundary condition at $r=1/q$, i.e. $\omega(1/q)=0$. 
This does not affect the final result, due to a very weak dependence of the integral~(\ref{omega_weak})  on the 
upper bound.
In this case, we obtain the following corrections in the region $r>R$ outside the star:
\begin{equation*}
\begin{split}
\frac{r^3\omega}{2 G \tilde{J}}-1 &=  12\varepsilon\left[\frac{a_3}{s_3(s_3-3)}\left(\frac{r}{r_V}\right)^{3}\left(\frac{r_V}{r_3}\right)^{s_3}+\frac{a_2}{s_2(s_2-3)}\left(\frac{r}{r_2}\right)^{s_2}\left(1-\left(\frac{r}{r_V}\right)^{3-s_2}\right)\right]H_{R< r\leq r_V}\\
&+\frac{12a_3\varepsilon}{s_3(s_3-3)}\left(\frac{r}{r_3}\right)^{s_3}H_{r>r_V}\; .
\end{split}
\end{equation*}
Assuming $s_2 <3$, one can write the solution in the regions $R<r\ll r_V$ and $r\gg r_V$ that we will focus on in the following:
\begin{equation}
\label{omega1sol_out}
\omega = \frac{2 G \tilde{J}}{r^3}\left[1 +\varepsilon\frac{12 a_2\varepsilon}{s_2(s_2-3)}\left(\frac{r}{r_2}\right)^{s_2}H_{R<r\ll r_V} + \frac{12 a_3\varepsilon}{s_3(s_3-3)}\left(\frac{r}{r_3}\right)^{s_3}H_{r\gg r_V}\right]\; .
\end{equation}
The above expression tells us how the corrections to $\omega$ outside the star can be read off from the coefficient $K_1$.

\paragraph{Inside the source:}
As we saw in the above subsection, the leading term differs from GR inside the star when $\kappa_2\neq-2$, meaning that the Vainshtein screening is broken. In theories for which $\kappa_2=-2$, the leading term in the solution for $\omega$ coincides with its GR counterpart, and the corrections to the frame-dragging function come from the subleading terms.
Assuming for simplicity that the star has a constant density $\rho_0$, the frame-dragging function inside the star can be written as follows:
\begin{equation}
\omega(r)-\omega(0) =- \frac{\rho_0\Omega r^2}{5 M_P^2}\left[1 + \frac{10 a_0\varepsilon}{(s_0+5)(s_0+2)}\left(\frac{r}{r_0}\right)^{s_0}- \frac{40 a_1\varepsilon}{(s_1+5)(s_1+2)}\left(\frac{r}{r_1}\right)^{s_1}\right]\; ,
\end{equation}
where $\omega(0)$ can be determined using Eq.~(\ref{omega1sol_out}) by continuity of $\omega$ at the surface of the star $r=R$. Once again, the subleading terms can be read off from the coefficients $K_1$ and $K_2$.

\section{Slow rotation in the weak-field approximation with a time-dependent scalar field}
\label{sec:weakfield1}

In this section, we study the slow rotation in DHOST Ia theories with $q\neq 0$, which means the scalar field is time dependent.
In addition to the weak-field assumption (\ref{cond1}), we also assume that 
\begin{equation}
\label{cond2}
\phi'^2\ll q^2 \; ,
\end{equation}
i.e. that the spatial gradient of the scalar field is much smaller than the time derivative of $\phi$. This can be viewed as a manifestation of the ``static'' Vainshtein screening and the failure of the Vainshtein mechanism for the time evolution of the scalar~\cite{Babichev:2011iz}.
Clearly, for static solutions, the condition~(\ref{cond2}) does not apply; therefore, we will not use it in the case of purely static configurations, see Sec.~\ref{sec:weakfield2}.
We will also assume that dimensionless combinations of coefficients are of $\mathcal{O}(1)$, for instance $ f\sim q^2 f_X\sim q^2 A_1\sim\mathcal{O}(1)$.
Under the assumptions~(\ref{cond1}) and (\ref{cond2}), the coefficients $K_1$ and $K_2$, Eqs.~(\ref{K1full}) and (\ref{K2full}), read
\begin{eqnarray}
\label{K1exp}
K_1 &=& \frac{4}{r}-\frac{\la'+\nu'}{2}+\frac{2(f_\p + q^2 A_{1\p})\p' -
\left(f_X+2A_1+q^2A_{1X}\right) \left( q^2 \nu' +2\phi'\phi'' \right)}{2(f+q^2 A_1)}\; ,\\
\label{K2exp}
K_2 &=& -\frac{1}{f+q^2 A_1}\left[1+\mathcal{O}\left(\la,\frac{\phi'^2}{q^2}\right)\right]\; ,
\end{eqnarray}
where we have used Eq.~(\ref{Xp}) in the weak-field approximation, and all the functions are evaluated at $\phi=qt$ and $X=q^2/2$.
The aim is to see how the solution to Eq.~(\ref{tpeq}) for $\omega$ is modified in the case of the scalar-tensor theories, with respect to the GR solution. 
We can see that generically the coefficient $\kappa_2$ defined in Eq.~(\ref{epsilon_coeffs}) is not the same as in GR, signaling that the screening is broken inside the source. If the condition $r\phi'\phi''/q^2\ll 1$ is verified, it is clear from Eq.~(\ref{K1exp}) that the corrections to $K_1$ are small compared to $4/r$, in which case $\omega$ has the same form as in GR at leading order outside the star, see section~\ref{sec:solutions}. For instance, this condition is satisfied if the solution for $\phi$ is a power law, and we will see in many examples below that this is generically the case.
Note that only the functions $f$ and $A_1$ directly appear in these coefficients. 
Of course the other functions of the Lagrangian enter the expression implicitly
via the scalar and metric functions in~(\ref{tpeq}).
However, we can immediately see that in a theory for which $f_X=A_1=0$ and the Vainshtein mechanism is effective in spherical symmetry, the coefficient $K_1$ is the same as in GR up to subleading corrections. 
Indeed, in this case we have
\begin{equation*}
K_1 =
\frac{4}{r}-\frac{\la'+\nu'}{2} + \frac{f_\phi}{f}\phi'\; .
\end{equation*}
When the Vainshtein mechanism in spherical symmetry is operational, the fifth force is screened for $r\ll r_V$, implying $\phi'\ll\{\la',\nu'\}$. Also, the solutions for $\{\la,\nu\} $ are those of GR at leading order. Assuming $f_\phi/f\lesssim\mathcal{O}(1)$, these two conditions show that 
the GR expression for $K_1$ is recovered up to $r_V$ suppressed corrections, which means that the subleading corrections for $\omega$ outside the star are also $r_V$ suppressed.

The Vainshtein mechanism in spherical symmetry was studied for DHOST Ia theories in \cite{Dima:2017pwp,Crisostomi:2019yfo,Hirano:2019scf,Langlois:2017dyl}. Adopting similar notations, we define
\begin{equation*}
x=\frac{\phi'}{r},\quad y = \frac{\nu'}{2r},\quad z = \frac{\la}{2r^2},\quad M(r)=4\pi \int_{0}^{r}\rho(\bar{r})\bar{r}^2\td\bar{r},\quad \ma(r)=\frac{G M(r)}{  q^2  r^3} \; .
\end{equation*}
Outside the source, we have $\ma = r_S/(2q^2 r^3)$, and we will define the Vainshtein radius $r_V$ as $\ma(r_V)\sim1$, meaning that
\begin{equation}
\label{rVdef}
r_V^3 \equiv \frac{r_S}{q^2}\; .
\end{equation}
The functions $\{\la,\nu\}$ vary slowly with time in this section, and we assume:
\begin{equation*}
\dot{z}\sim q z, \quad  \dot{y}\sim q y\; ,
\end{equation*}
which can be checked once the solutions for $\{y,z\}$ are found. 
The $(tt)$ and $(rr)$ field equations for the metric, Eqs.~(\ref{Emetric00}) and (\ref{Emetric11}), can be solved in terms of $x$ and $\ma$, and written in the form:
\begin{eqnarray}
\label{ysol}
y &= \alpha_1 \ma + \beta_1 x + \gamma_1 x^2 + \delta_1 r x x' + \eta_1\; ,\\
\label{zsol}
z &= \alpha_2 \ma + \beta_2 x + \gamma_2 x^2 + \delta_2 r x x'+ \eta_2\; ,
\end{eqnarray}
where all the time dependent coefficients are listed in Appendix~\ref{appendix:a}. They can be expressed in terms of the Lagrangian functions evaluated on the background $\phi= qt$ and $X=q^2/2$. Note that these coefficients are not necessarily dimensionless. 
In order to obtain the above equations~(\ref{ysol}) and~(\ref{zsol}), we have also assumed $r\ll 1/q$. 
In terms of the function $\ma$ defined above, the Vainshtein screening in the nonrotating case generally happens when $\ma\gg 1$, which corresponds to $r\ll r_V$. However, there are deviations from GR when $\ma\ll 1$, which we will examine in the region $r_V\ll r \ll 1/q$ where our equations are valid.
The terms we neglected should be kept if we want to match to the appropriate de Sitter solution at cosmological radii $r\geq 1/q$. This is the asymptotic condition consistent with the linear time dependence of the scalar field, as discussed in \cite{Babichev:2012re} for the cubic Galileon theory. 

The expressions~(\ref{ysol}) and (\ref{zsol}) for $y$ and $z$ can then be used in the scalar field equation, Eq.~(\ref{Ephi}), yielding a cubic equation for $x$ \cite{Dima:2017pwp}:
\begin{equation}
\label{scalarIa}
C_3 x^3 + C_2 x^2 + \left(C_1 + \Gamma_1 \ma + \Gamma_2 \frac{(r^3\ma)'}{r^2}\right)x + \Gamma_0 \ma + \eta_3 = 0\; .
\end{equation}
Substituting equations (\ref{ysol}) and (\ref{zsol}) in~(\ref{K1exp}) results in:
\begin{equation}
\label{K1Ia}
K_1 = \frac{4}{r}\left[1 + \alpha_0 r^2 \ma + \zeta_0 \left(r^3 \ma\right)'+\beta_0 r^2x + \kappa_0 r^3 x'+ \gamma_0 r^2 x^2 + \delta_0 r^3 x x' + \sigma_0 r^4\left(xx'' + x'^2\right) + \eta_0 r^2 \right]\; .
\end{equation} 
The coefficients of~(\ref{scalarIa}) and~(\ref{K1Ia})  are listed in Appendix~\ref{appendix:a}.
One can see from Eq.~(\ref{K1Ia}) that there is always a leading term in the brackets corresponding to the Minkowski limit of the metric  $K_1 \simeq 4/r$ (for radii $ r \ll 1/q$).
We discuss below various cases of Eq.~(\ref{scalarIa}) leading to different nonrotating solutions \cite{Dima:2017pwp,Kimura:2011dc}.
Substituting the relevant solution for $x$ in Eq.~(\ref{K1Ia}), we will examine how the modification of gravity affects slowly rotating solutions, i.e. the equation (\ref{omegaeq_weak}) for $\omega$. 
We will show that the leading corrections to the coefficients $K_1$ and $K_2$ are small and take the form of power laws. In this case, we showed in section~\ref{sec:solutions} that $\omega$ has the GR form at leading order outside the star, up to an overall factor 
(which can be absorbed in the definition of the angular momentum of the star as measured by an exterior observer, unless the density distribution of the  star is known). On the other hand, the screening can be broken inside the star. 
We will be interested in the subleading corrections to $\omega$ when the leading term is not modified and compare them to those of GR.
\subsection{Outside the Vainshtein radius}
We first examine the linear regime outside the Vainshtein radius, where we have $\ma\ll 1$. There are two different cases, depending on the coefficient $\eta_3$. In this regime the Vainshtein mechanism for nonrotating sources does not operate, and the corrections to the metric for the spherically symmetric solution are expected to be large. Therefore, we also expect that the equation for $\omega$ receives corrections larger than those inside the Vainshtein radius. 

\subsubsection{\texorpdfstring{$\eta_3=0$}{eta30} and \texorpdfstring{$C_1\neq 0$}{C10}}
Let us first consider the case $\eta_3=0$. A sufficient condition for this coefficient to vanish is $K = G_{3\phi} = 0$. 
In this case, the nonlinear terms in $x$ in Eq.~(\ref{scalarIa}) can be neglected, and the solution for $x$ is
\begin{equation*}
x = -\frac{\Gamma_0}{C_1}\ma\sim \frac{r_S}{r^3}\; .
\end{equation*}
Substituting this expression into Eq.~(\ref{K1Ia}), we obtain the expression for $K_1$,
\begin{equation}
\label{K1lin1}
K_1 = \frac{4}{r}\left[1 +\mathcal{O}\left(\frac{r_S}{r}\right)\right]\; .
\end{equation}
This shows that the corrections due to the scalar field are not suppressed by powers of the Vainshtein radius, and are of the order of the Newtonian potentials. This is expected in the region $r\gg r_V$ where the Vainshtein screening in spherical symmetry is no longer effective (meaning we do not have $\la'+\nu'\simeq0$ in Eq.(\ref{K1exp})).

\subsubsection{\texorpdfstring{$\eta_3\neq 0$}{etaneq0} and \texorpdfstring{$C_1\neq 0$}{C1neq0}}
If $\eta_3\neq 0$, we have $\Gamma_0 \ma\ll \eta_3$, since $\mathcal{A}\ll 1$. In this case Eq.~(\ref{scalarIa}) reduces to the following cubic equation for $x$ with $r$-independent coefficients:
\begin{equation*}
C_3 x^3 + C_2 x^2 + C_1 x + \eta_3 = 0\; .
\end{equation*}
The relevant solution for $x$ must be chosen by taking into account the asymptotic behavior of the solution at large radii, $r\gg1/q$. Since the coefficients of the algebraic equation depend only on time, $x$ does not depend on the radial coordinate and we have $x = x_0(t)$. Substituting this solution into Eq.~(\ref{K1Ia}), we obtain
\begin{equation}
\label{K1lin2}
K_1 = \frac{4}{r}\left[1 + \mathcal{O}\left(q^2r^2\right)\right]\; .
\end{equation}
Note that here the corrections have a clear physical interpretation; they arise as a backreaction on the metric due to the ``weight'' of the scalar field, see e.g.~\cite{Babichev:2018rfj}. They are present even in the simplest theory with a minimally coupled scalar field.
The corrections are larger in this case than for $\eta_3=0$, considered above. 
Indeed, using Eq.~(\ref{rVdef}), we obtain that the ratio of the corrections in~(\ref{K1lin1}) to the corrections in~(\ref{K1lin2}) are of order
$\left(r_V/r\right)^3$. 

In the rest of this section, we will consider the region $ r\ll r_V$, where the Vainshtein mechanism usually operates in spherical symmetry.

\subsection{Case 1: \texorpdfstring{$C_3\neq0$}{C3} and \texorpdfstring{$\Gamma_1\neq0$}{C1}, inside the Vainshtein radius}
We first consider the generic case $\Gamma_1 \neq 0$ and $C_3\neq 0$ (see Appendix~\ref{appendix:a} for their expressions).
Note that when $\Gamma_1=0$ then we also have $C_3=0$. 
We assume that $C_3\Gamma_1<0$ and we will confirm this choice later. 
Then the solutions to (\ref{scalarIa}) for $r\ll r_V$ are\footnote{
For some theories these solutions have been shown to match a de~Sitter asymptotic~\cite{Babichev:2016jom}.},
\begin{equation}
\label{x1V}
x_1 = \pm\sqrt{\frac{-\Gamma_1 \ma- \Gamma_2 \frac{(r^3\ma)'}{r^2}}{C_3}}\; ,
\end{equation}
where we used $\ma\gg 1$ to simplify. 
The $\pm$ sign must be chosen in order to match the solution at infinity, depending on the theory. Outside the star $(r^3\ma)'=0$; therefore, our choice  $C_3\Gamma_1<0$ is indeed correct to have a real solution in the exterior region. 
Extra conditions should be also imposed on $\Gamma_2$ for the argument of the square root to be positive. In particular, a sufficient condition is $\Gamma_2<0$. 
We do not consider the third solution to the cubic equation, since there is no known example where it is matched to de Sitter asymptotics. (Note however that in~\cite{Kimura:2011dc} the asymptotically flat spherically symmetric solutions of this branch were found, and it was shown that the Vainshtein mechanism is not effective for this branch unless the speed of gravitational waves $c_T=1$). Substituting the solution~(\ref{x1V}) for $x$ in (\ref{K1Ia}), we obtain
\begin{equation}
\label{K1x1}
K_1 = \frac{4}{r}\left[1 + \frac{\td}{\td r}(\iota_0 r^3\ma + \iota_1 r^4 \ma' + \iota_2 r^5 \ma'') 
+ \mathcal{O}\left(q^2r^2\sqrt{\ma}\right)\right]\; ,
\end{equation}
where we assumed $\ma\sim r^n\ma^{(n)}$ for the subsubleading part, and the expressions for the $\iota_i$ are listed in Appendix~\ref{appendix:a}. The above coefficient $K_1$ generically differs from its GR counterpart inside the source (see Eq.~(\ref{omegaweakGR})).
In particular, as can be seen from Eq.~(\ref{omegaweakGR}), $\iota_1 = \iota_2 = 0$ in GR.
In the exterior region outside the star, $R <r \ll r_V$, we have $(r^3 \ma)' =0$ and the previous equation simplifies to
\begin{equation*}
 K_1 = \frac{4}{r}\left[1 +  \mathcal{O}\left(\frac{r_S \sqrt{r}}{r_V^{3/2}}\right)\right]\; .
\end{equation*}
The corrections to the solution for $\omega$ are subdominant, as we showed in section~\ref{sec:solutions}. Furthermore, they are suppressed by powers of $r_V$, in an analogous way to the screening in spherical symmetry. In fact, the screening is even more effective for $\omega$, since one has a power $r_S/r_V$ instead of $r/r_V$ as in Eq.~(\ref{metric_suppression}).
A similar screening also happens for the third solution to (\ref{scalarIa}) that cannot be matched to the de Sitter solution at large radii, which we do not consider here.

\subsubsection{A class of shift symmetric beyond Horndeski theories}
Let us now restrict ourselves to the quadratic sector of GLPV theories \cite{Gleyzes:2014dya}, which corresponds to the Lagrangian~(\ref{BHtheory}).
In the case of shift-symmetric beyond Horndeski theories, the Vainshtein mechanism for spherically symmetric configurations has been extensively studied. In particular, 
in Ref.~\cite{Babichev:2016kdt} it was shown that the backreaction of the scalar field on the metric leads to a redefinition of Newton's constant $G$. 
Also, in a subclass of the theory, the Vainshtein screening has been considered for slowly rotating sources.
Indeed, the specific case of constant $A_3$ was studied in \cite{Babichev:2016jom,Sakstein:2016oel} for relativistic stars. It was shown in this theory that $\omega$ satisfies the GR equation outside the star, meaning that $K_1 = 4/r$ exactly, with no subleading corrections. This result remains true for the shift symmetric theories defined above, and does not rely on the weak-field approximation, as we discussed in section~\ref{sec:relativistic}. 

Here we discuss the equation for $\omega$ in the weak-field approximation inside the matter source. 
After substituting the solution for $x$, given in~(\ref{x1V}), the metric potentials read
\begin{equation}
\label{metric_BHshift}
\begin{split}
y &= \tilde{G}\left(\frac{M}{r^3} - \frac{q^4 A_3^2}{2[f(q^2 A_{3X}+ 4 A_3 + 2f_{XX})+ q^2 A_3 f_X + 2f_X^2 ]}\cdot\frac{M''}{r}\right)\; ,\\
z &=  \tilde{G}\left(\frac{M}{r^3} + \frac{q^2A_3(q^4 A_{3X} +2f_X + 5q^2 A_3 +2 q^2 f_{XX})}{2[f(q^2 A_{3X}+ 4 A_3 + 2f_{XX})+ q^2 A_3 f_X + 2f_X^2 ]}\cdot\frac{M'}{r^2}\right)\; ,
\end{split}
\end{equation} 
where we have defined an effective gravitational  constant:
\begin{equation*}
\tilde{G}= \frac{G}{2f- 4 q^2 f_{X} - 2q^4 f_{XX}-5q^4 A_3 - q^6 A_{3X} }\; .
\end{equation*}
The above equations show that the Vainshtein mechanism in spherical symmetry is broken inside the source~\cite{Kobayashi:2014ida},
but that GR is recovered in the exterior region where $M$ is constant.

Substituting the metric potentials in Eq.~(\ref{tpeq}) with the coefficients~(\ref{K2exp}) and (\ref{K1Ia}), the equation for $\omega$ inside the star and in the weak-field limit reads
 \begin{equation*}
\omega'' + \frac{4}{r}\left[1 -\frac{G M'}{4(2f -2q^2 f_{X} -  q^4 A_3)}\right]\omega'=- \frac{4 G M'\Omega}{r^2 (2 f -2q^2 f_X -  q^4 A_3)}\left[1+\mathcal{O}(r^2z)\right] \; ,
 \end{equation*}
which is the same equation as in GR (up to the subleading term in the coefficient $K_2$) provided we redefine Newton's constant as:
\begin{equation*}
G^* =  \frac{G}{2f - 2q^2 f_X -  q^4 A_3}\neq \tilde{G}\; .
\end{equation*}
As we can see, in general the two redefined Newton constants $\tilde{G}$ and $G^*$ do not coincide. 
This means that the coefficient $\kappa_2$ is not the same as in GR, and the Vainshtein screening is broken inside the star (see Eq.~(\ref{omega_in_leading})). This is expected for $A_3\neq0$ since the Vainshtein screening for static sources is broken inside matter for these theories \cite{Kobayashi:2014ida}. However, this remains true even for $A_3=0$ when the Vainshtein screening in the nonrotating case works inside the star (as can be seen from Eq.~(\ref{metric_BHshift})). The two redefinitions of $G$ coincide in theories with $f_X=A_3=0$, but in this case $\Gamma_1=C_3=0$, so the analysis of the present section is not valid.

\subsection{Case 2: \texorpdfstring{$C_3=\Gamma_1=0$}{C3G10} and \texorpdfstring{$C_2\neq0$}{C2neq0} inside the Vainshtein radius}
In this section we consider a particular case of the DHOST Ia theories,
\begin{equation}
\label{case2cond}
f A_{1X} + A_1 f_X - f A_3 = 0\; ,
\end{equation}
which implies $C_3=\Gamma_1=0$.
In this case~(\ref{scalarIa}) is quadratic, and the general solution is
\begin{equation}
\label{x2exp}
x_2 = -\frac{r^2C_1 + \Gamma_2(r^3 \ma)' \pm\sqrt{[r^2C_1 + \Gamma_2(r^3 \ma)']^2 - 4r^4\ma C_2\Gamma_0} }{2r^2C_2}\; .
\end{equation}
 Assuming $\Gamma_0 C_2 < 0$ and neglecting $C_1$ in the limit $\mathcal{A}\ll 1$, the solution for $x_2$ in the exterior region $R< r \ll r_V$ reads
\begin{equation}
\label{x2out}
x_{2\text{out}} = \pm\sqrt{\frac{-\ma\Gamma_0}{C_2}}\; .
\end{equation}
Substituting this expression in (\ref{K1Ia}), we obtain
\begin{equation}
\label{K1out2}
K_1 = \frac{4}{r}\left[1 + \xi \frac{r_S}{r} + \mathcal{O}\left(\frac{r_S \sqrt{r}}{r_V^{3/2}}\right)\right]\; ,
\end{equation}
where the coefficient $\xi$ reads
\begin{equation*}
\xi = \frac{\alpha_0}{2}+ \frac{\Gamma_0}{4 C_2}\left[3\left(\delta_0 - 4\sigma_0\right)-2\gamma_0\right]\; .
\end{equation*}
The full expression is rather lengthy, but it can be rewritten in the form
\begin{equation}
\label{xiexp}
\xi = \left[f+q^2A_1\right]\left[f\left(q^2 A_{1X} + 2 f_X\right) + A_1\left(2f + q^2 f_X\right)\right] \xi_0 \; ,
\end{equation}
where the coefficients $\xi_0$ is in general a time dependent function. 

Several interesting observations can be made from~(\ref{xiexp}). First of all, 
for theories with $A_1=A_3=f_X=0$ (we also used the condition~(\ref{case2cond})), one automatically obtains $\xi=0$. 
This means that there are only subleading (Vainshtein suppressed) corrections to the coefficient $K_1$, see Eq.~(\ref{K1out2}). 
For example, this is the case for the cubic Galileon, which we will discuss in more detail below. 

In fact, from~(\ref{K1out2}) one can draw a conclusion for more general theories, namely those satisfying~(\ref{case2cond}) with $f_X\neq 0$. Indeed, assuming that the Vainshtein mechanism in spherical symmetry is at work, the metric potentials approximately verify the GR relation in vacuum:
\begin{equation*}
y-z  = 0 \; . 
\end{equation*}
 This relation is valid whenever the Vainshtein mechanism in spherical symmetry operates outside the star, up to subleading corrections.
Substituting Eq.~(\ref{x2out}) into Eqs.~(\ref{ysol}) and (\ref{zsol}), we obtain
\begin{equation}
\label{static_cond}
	r^2 \left(y-z\right) =  -4\xi_0f_X\left(f+ q^2 A_1\right)^2  \cdot\frac{r_S}{r} + \mathcal{O}\left(\frac{r_S\sqrt{r}}{ r_V^{3/2}}\right)\; .
\end{equation}
We see that if $f_X\neq0$, one must impose $\xi_0=0$ to recover the Vainshtein screening in the absence of rotation. In this case, $\xi$ also vanishes, see  Eq.~(\ref{xiexp}). 
On the other hand, when~(\ref{case2cond}) is satisfied but $f_X=0$, the Vainshtein mechanism operates in spherical symmetry, but $\xi$ is not necessarily zero. We will consider an explicit example below.

\subsubsection{Example 1: Theory with larger corrections to the frame-dragging function in the exterior region}
As we mentioned above, there are theories which allow for the spherically symmetric Vainshtein screening, 
but for which the corrections to the frame-dragging equation are of the order of the Newtonian potentials (showing that the screening is less effective for $\omega$ than for the metric potentials).
We consider such theories in detail in the present section. For theories verifying \ref{case2cond}, a necessary condition for the Vainshtein mechanism to work in spherical symmetry if $\xi_0\neq0$ is $f_X=0$. This can be seen from Eq.~(\ref{static_cond}), which shows deviations of the metric functions from the GR case.
If we assume $f_X=0$, Eq.~(\ref{case2cond}) implies $A_3 = A_{1X}$. 
For simplicity, we will set $f=1/2$, although the result below can be extended for a generic $f$.
In this case, the dimensionless coefficient $\xi$ reads
\begin{equation*}
\xi = - \frac{\left(1 + 2 q^2A_1\right)\left(2 q^2A_1 + q^4 A_{1X}\right)\left[2 A_{1\phi}\left(4 + 6q^2 A_1 - q^4 A_{1X}\right) + \left(1 + 2 q^2 A_1\right)\left(3 G_{3X} + 2 q^2 A_{1\phi X}\right)\right]}{2\left(2 + 6 q^2 A_1 + q^4 A_{1X}\right)^2\left[3 A_{1\phi}\left(2 + 2q^2 A_1 -  q^4 A_{1X}\right)+2\left(1 + 2 q^2 A_1\right)\left( G_{3X} +  q^2 A_{1\phi X}\right)\right]}\; .
\end{equation*}
Note that since we consider the case $C_2\neq 0$, either  $G_{3X}\neq0$ and/or $A_{1\phi}\neq0$. The metric potentials in these theories read:
\begin{align}
\label{ycex}
r^2 y &= \iota_3\frac{r_S}{r} + \mathcal{O}\left(\frac{r_S \sqrt{r}}{r_V^{3/2}}\right)\; ,\\
\label{zcex}
r^2 z &= \iota_3\frac{r_S}{r} + \mathcal{O}\left(\frac{r_S \sqrt{r}}{r_V^{3/2}}\right)\; ,
\end{align}
where the coefficient $\iota_3$ is given in Appendix~\ref{appendix:a}.

After redefining Newton's constant, the metric potetentials have the GR form up to subleading corrections, meaning that the Vainshtein mechanism works in spherical symmetry.
Meanwhile $\xi\neq 0$, and therefore the corrections to the frame-dragging function $\omega$ are of order $r_S/r$, as Eq.~(\ref{K1out2}) shows.  
This implies that the screening for $\omega$ is not as effective as it is for the metric potentials $\lambda$ and $\nu$. We have thus demonstrated for a particular theory that the Vainshtein screening in spherical symmetry is not sufficient to ensure that the leading corrections to the GR expression for $\omega$ are suppressed by powers of $r_V$.

\subsubsection{Example 2: Theory with \texorpdfstring{$c_T=1$}{cT1} and no decay of the graviton into dark energy}
Most of the DHOST Ia theories as models of dark energy \cite{Langlois:2017mxy} have been ruled out by the constraint $c_T=1$ (the graviton propagates at the speed of light) coming from the merger of a binary neutron star system \cite{TheLIGOScientific:2017qsa,Creminelli:2017sry,Ezquiaga:2017ekz}, and requiring that the graviton does not decay into dark energy \cite{Creminelli:2018xsv}. 
The surviving theories correspond to the choice
\begin{equation}
\label{survivor}
2f A_4=3 f_X^2\; , \quad A_1=A_2=A_3=A_5=0\; .
\end{equation}
The Vainshtein screening  in the absence of rotation for these theories was studied in \cite{Crisostomi:2019yfo,Hirano:2019scf}. It was shown that the screening is broken inside the star, and that it may work in the exterior region provided the parameters of the theory are fine-tuned.

\paragraph{Outside the source:} Outside the star, the coefficient $K_1$ is of the form (\ref{K1out2}), with
\begin{equation*}
\xi = \frac{f_X[2f_\phi(f+ 5 q^2 f_X)-3 f q^2 G_{3X} - 2 f q^2 f_{X\phi}]}{8(q^2 f_X - 2f)^2(fG_{3X} - 3 f_X f_\phi)}\; ,
\end{equation*}
where we used~(\ref{survivor}).
Note that the denominator does not vanish in the case $C_2\neq0$.
It was shown in \cite{Hirano:2019scf,Crisostomi:2019yfo} that the Vainshtein mechanism can work outside the star in this theory if the parameters verify
\begin{equation}
\label{survivor_cond_out}
f_X\left[2f_\phi(f+ 5 q^2 f_X)-3 f q^2 G_{3X} - 2 f q^2 f_{X\phi}\right]=0\; ,
\end{equation}
which is exactly the condition for $\xi$ to vanish, as can be seen from the above expression. This shows that if we fine-tune the parameters to recover the Newtonian potentials outside the source, then the screening for $\omega$ becomes more effective, in the sense that corrections to the GR expression for $\omega$ are suppressed by powers of $r_V$ (see Eq.~(\ref{K1out2})).

\paragraph{Inside the source:}
Let us examine the $(t\vp)$ equation inside the source, where $(r^3 \ma)'\neq 0$ and we assume $r \ma'\sim \ma$. We also assume
\begin{equation*}
\Gamma_2 =  24 q^2 f^2\left(2f - q^2 f_X\right)  f_X\neq 0 \; ,
\end{equation*} 
which implies that the leading term inside the square root of (\ref{x2exp}) is the one containing $\Gamma_2$. One of the branches obtained with these assumptions is physically unacceptable, as argued in~\cite{Hirano:2019scf}, so we focus on the second branch for which
\begin{equation*}
r^2x \simeq -\frac{\Gamma_0}{\Gamma_2}\frac{r^4 \ma}{(r^3 \ma)'}\sim\mathcal{O}\left(q^2r^2\right)\; .
\end{equation*}
This expression is only valid when $(r^3 \ma)'\neq 0$ and $(r^3 \ma)'\gg1$. 
We assume that the Vainshtein mechanism in spherical symmetry operates outside the source, meaning that condition (\ref{survivor_cond_out}) is verified. Additionally, one must rescale Newton's constant as $\tilde{G}= G/(2f-q^2f_X)$. 
The frame-dragging equation inside the star reads
\begin{equation*}
\omega'' +\frac{4}{r}\left[1-\frac{q^2 f_X}{2f-q^2 f_X}\frac{ \tilde{G}M}{r}-\frac{ (f-q^2 f_X)\tilde{G}M'}{2(2f-q^2 f_X)} + \mathcal{O}\left(q^2 r^2\right)\right]\omega'=-\frac{2 (2f-q^2 f_X)\tilde{G}M'\Omega}{f r^2}\left[1+\mathcal{O}(r^2z)\right]\; .
\end{equation*}
On the left-hand side of this equation there is an extra term $\propto \tilde{G}M/r$ compared to the equation in GR (see Eq.~(\ref{omegaweakGR})).
Note that this term is nonzero, since we study the case $\Gamma_2\neq0$, which implies $ f_X \neq 0$ (see Appendix~\ref{appendix:a}). 
The screening is broken inside the star, because generically $\kappa_2\neq -2$ when $f_X\neq0$ as can be seen from the equation above. This behavior is not surprising, since the Vainshtein mechanism in spherical symmetry is broken inside the source.
Note that the expressions for $K_1$ inside and outside the star were obtained in different limits ($(r^3 \ma)'\gg 1$ in the former and $(r^3 \ma)'=0$ in the latter case), therefore they cannot be matched at the surface of the star. One would have to solve the full equation to obtain a continuous profile, as was done in Ref.~\cite{Hirano:2019scf}.

\subsubsection{Example 3: Cubic Galileon}
The time-dependent cubic Galileon was studied in Ref.~\cite{Kimura:2011dc}, and also in Ref.~\cite{Babichev:2012re}, where the appropriate de Sitter asymptotics were discussed. To get the cubic Galileon from the general action~(\ref{actionDHOST}), we set
\begin{equation}
\label{CG}
f = \frac{1}{2},\quad  G_3 = \beta X, \quad K = \eta X, \quad A_i = 0\; .
\end{equation}
With these choices, the coefficient $\Gamma_2 = 0$ in equation~(\ref{scalarIa}), and the expression for $x$ for $r\ll r_V$ reads
\begin{equation*}
x= \pm q^2\sqrt{\frac{\ma}{2}}\; .
\end{equation*}
The sign should be chosen when properly examining the asymptotic behavior for large radii, but it does not affect the resulting equation for $\omega$ (since quadratic terms are dominant in (\ref{K1Ia}) inside the Vainshtein radius)
both inside and outside the star. Then the equation for the frame-dragging function can be written as follows:
\begin{equation*}
\omega'' +\frac{4}{r}\left[1-\frac{GM'}{4} + \mathcal{O}\left(q^2r^2\sqrt{\ma}\right)\right]\omega'=-\frac{4 GM'\Omega}{ r^2}\left[1 + \frac{2GM}{r} + \mathcal{O}\left(q^2r^2\sqrt{\ma}\right)\right]\; ,
\end{equation*}
where we have assumed $\ma\sim r\ma'$ and $\beta q^2 \sim 1$. 
By comparing  the above equation with Eq.~(\ref{omegaweakGR}) and taking into account~(\ref{rVdef}), 
we can see that the corrections for $\omega$  to the GR equation are suppressed by powers of $r_V$ inside the Vainshtein radius both inside and outside the source. Using the results of section~\ref{sec:solutions}, we then conclude that deviations from the GR expression for $\omega$ are also suppressed by powers of $r_V$ in a way analogous to the screening in spherical symmetry. It should also be noted that nonlinear GR corrections (which we did not take into account) may be larger than those due to modified gravity, but they are of course still smaller than the linear GR terms.

\section{Slow rotation in the weak-field approximation with a static scalar field}
\label{sec:weakfield2}

In this section, we will set $q=0$, meaning $\phi = \phi(r)$, and consider the shift-symmetric sector of the DHOST Ia class with an additional linear coupling of the scalar field to the Ricci scalar of the form $\alpha\phi R$. This term breaks the shift symmetry, i.e. $\phi\to \phi+ $const is no longer a symmetry of the action, and allows one to escape the no hair-theorem of Ref.~\cite{Lehebel:2017fag}.
The coupling to the Ricci scalar provides a nontrivial scalar field configuration with rich phenomenology, including k-mouflage gravity~\cite{Babichev:2009ee}---an analog of the Vainshtein mechanism. In this setup, the scalar equation can be written in the form
\begin{equation*}
\nabla_\mu J^\mu = -\alpha R \; ,
\end{equation*}
where $J^\mu = -\frac{\delta\mathcal{L}}{\delta(\del_\mu\phi)}$ is the conserved current associated with the shift symmetry of the action when $\alpha=0$. 
An interesting property of the above equation is that it can be integrated in the weak-field regime:
\begin{equation*}
\frac{1}{r^2}\frac{\td}{\td r}\left[r^2 J^r + \alpha\left(2r\la-r^2\n'\right)\right]=0\; .
\end{equation*}
Even though the action is not shift-symmetric if $\alpha\neq 0$, there is an effective conserved current in the weak-field limit in this particular case of a linear coupling to the Ricci scalar. In the following, we will set the integration constant to 0, in order for the norm of the current $J_\mu J^\mu=e^\la(J^r)^2$ to be regular at the center of the star. In this case, the scalar field equation reads
\begin{equation}
\label{scalareq_alpha}
r J^r + \alpha\left(2\la-r\n'\right)=0\; .
\end{equation}
As it is clear from the above equation, the radial component of the current cannot be zero, in contrast to the shift-symmetric time-dependent case, 
where we have $J^r =0$. The presence of the symmetry-breaking term renders $J^r$ nonzero. 
This, and the fact that the ratio $\phi'^2/q^2$ is no longer a small parameter in the equations, cf. Eq.(\ref{cond2}), 
changes the results in the weak-field approximation. This means that one cannot simply set $q=0$ in the above study, Sec.~\ref{sec:weakfield1}, and we have to proceed starting from square one. 

\subsection{K-essence}
\label{seckessence}
Let us first consider a k-essence theory, i.e. we take $f=1/2$, $A_i =0$ and $K = \mu X^p$, where $p\in\mathbb{N}\backslash \{0,1\}$ and $\mu$ is constant. Neglecting the backreaction of the energy-momentum of the scalar field on the metric, which corresponds to neglecting nonlinear scalar contributions in Eqs.~(\ref{Emetric00}) and (\ref{Emetric11}), one can integrate Eq.~(\ref{Emetric00}) to obtain the metric potentials:
\begin{equation}
\label{kessence}
\begin{split}
\la &= \frac{2 G M(r)}{r}+2 \alpha r \p'\; ,\\
\nu' &=\frac{2G M(r)}{r^2}- 2\alpha\p'\; .
\end{split}
\end{equation}
After combining these expressions with Eq.~(\ref{kessence}), we obtain the following equation for the scalar field:
\begin{equation}
\label{scalarKE}
\frac{\alpha GM(r)}{r^2}+3\alpha^2\phi'-p\left(-\frac{1}{2}\right)^p\mu\p'^{2p-1}=0\; .
\end{equation}
Eq.~(\ref{scalarKE}) corresponds to the weak-field limit of the scalar equation in the Einstein frame, obtained by rewriting the action in terms of the conformal metric $\bar{g}_{\mu\nu} = e^{2\alpha\phi} g_{\mu\nu}$. In the weak-field limit, the conformal transformation corresponds to the change $\la\to \bar{\la}+2\alpha r \phi'$ and $\nu\to\bar{\nu}-2\alpha\phi$. 
In this frame, $\{\bar{\nu},\bar{\la}\}$ satisfy Eqs.~(\ref{kessence}) with $\phi=0$. The conformal transformation introduces a coupling of $\phi$ to matter, and a kinetic term for the scalar field, which explains the appearance of the first two terms in Eq.~(\ref{scalarKE}).

\subsubsection{Linear regime and Vainshtein radius}
Outside the star, we have $2GM(r)=r_S$, and in the limit $r\to\infty$ we can neglect the nonlinear term in~(\ref{scalarKE}). 
Then the solution for the scalar field  can easily be found:
\begin{equation}
\label{philin}
\phi' = -\frac{r_S}{6\alpha r^2}\; .
\end{equation}
Note that the limit $\alpha\to 0$ is not well defined in~(\ref{philin}). This is a consequence of the absence of a standard kinetic term in the considered theory. Indeed, due to the mixing term $\alpha\phi R$ the scalar degree of freedom has a kinetic term; however, it disappears in the limit $\alpha\to 0$, thus making the theory strongly coupled in this limit. 
Said differently, the nonlinear term is dominant for small $\alpha$, therefore the linear regime is nowhere valid.

Using~(\ref{philin}) in the first two equations of~(\ref{kessence}), one can see that in the linear regime, the GR condition $\la+\nu=0$ is not satisfied even approximately. Instead, the solution of (\ref{kessence}) reads
\begin{equation}
\label{metriclin}
\la = \frac{2 r_S}{3 r}\; , \quad \nu = -\frac{4 r_S}{3 r}\; ,
\end{equation}
meaning that deviations of the Newtonian potentials from the GR solutions are of $\mathcal{O}(1)$. Upon substituting~(\ref{philin}) and~(\ref{metriclin}) into Eq.~(\ref{K1exp}) for $\omega$, we obtain
\begin{equation}
\label{omegaeqk}
\omega''+\frac{4}{r}\left(1-\frac{r_S}{6r} \right)\omega'=0 \; .
\end{equation}
This expression is to be compared  with~(\ref{omegaweakGR}) in vacuum, for which $M'(r)=0$. 
One can see that the term proportional to $r_S$ has a coefficient different from the GR case.
 Thus, according to the results of section (\ref{sec:solutions}), the leading term in the solution for $\omega$ is the same as in GR, unlike the metric potentials. However, the subleading corrections in the weak-field approximation are of order $r_S/r$, meaning that the screening is less effective than in the region $r\ll r_V$, where the leading corrections are suppressed by powers of $r_V$.

\subsubsection{Inside the Vainshtein radius}
The linear regime breaks down at the Vainshtein radius $r\sim r_V$ where nonlinear terms become important.
Let us determine the Vainshtein radius by taking the solution for $\phi'$ at infinity and evaluating at which radius the nonlinear term becomes comparable to the linear one~\cite{Babichev:2009ee}. We find
\begin{equation}
\label{rVK}
r_V^2\sim\frac{r_S}{6}\left(\frac{|\mu| p}{3\cdot 2^p \alpha^{2p} }\right)^\frac{1}{2p-2}\; .
\end{equation} 
For $ r \ll r_V$, we can neglect the linear term in Eq.~(\ref{scalarKE}), and in this range of radii the scalar field reads
\begin{equation}
\label{kVsol}
\phi' = \text{sgn}\left[(-1)^p\alpha\mu\right]\left(\frac{2^p|\alpha| G M(r)}{p|\mu| r^2}\right)^{\frac{1}{2p-1}}\; .
\end{equation}
Note that the limit $\alpha\to 0$ is well defined, and we have $\phi'\to 0$, in contrast to the solution in the linear regime Eq.~(\ref{philin}). 
The limit is consistent with the solution to the scalar equation~(\ref{scalarKE}) for $\alpha=0$.
In the limit $\alpha \to 0$ the Vainshtein radius~(\ref{rVK}) is infinite; therefore, the Vainshtein mechanism operates for all distances and the linear regime is invalid. 

In the region $r\ll r_V$, one can compare the strength of the fifth force {with the Newtonian force 
$\{ \la'_{\text{GR}},\nu'_{\text{GR}} \}~\sim~2GM/r^2$, obtained by setting $\phi=0$ in (\ref{kessence}):
\begin{equation*}
	\left|\frac{\phi'}{\{\la'_{\text{GR}},\nu'_{\text{GR}}\}}\right|\sim\frac{1}{6\alpha}\left(\frac{r_S r^2}{2GM(r)r_V^2 }\right)^{\frac{2p-2}{2p-1}}\; .
\end{equation*}
Outside the source, in the region $R\leq r\ll r_V$, we have $2GM = r_S$, and it is clear that the fifth force is screened. Inside the star, assuming it has a constant density $\rho_0$, we have $2GM(r) = r_S r^3/R^3$. In this case, it is clear from the above expression that the fifth force becomes dominant for radii smaller than $r_* = R^3/r_V^2\ll R$. Meanwhile, in the region $r_*\ll r\leq R$, the fifth force is screened.
To examine the effects of rotation, we substitute Eq.~(\ref{scalarKE}) into the $(t\vp)$ metric equation. Assuming for instance that $(-1)^p\alpha\mu>0$ (the other case is analogous), the coefficients $K_1$ (outside and inside the star respectively) and $K_2$ read
\begin{align*}
	K_1^\text{out}&=\frac{4}{r}\left[1 + \mathcal{O}\left(\frac{r_S}{r_V}\left(\frac{r}{r_V}\right)^{\frac{2p-3}{2p-1}}\right)\right]\; ,\\
	K_1^\text{in}&=\frac{4}{r}\left[1 + \frac{r_S r}{8 r_V^2 }\left(\frac{3-4p}{3(1-2p)}\left(\frac{r}{r_*}\right)^\frac{1}{2p-1}-\frac{3r}{r_*}\right)\right]\; ,\\
K_2 &= -2 \left[1 + \frac{r_S r^2}{R^3}\left(1 + \mathcal{O}\left(\frac{r_*}{r}\right)^\frac{2p-2}{2p-1}\right)\right]\; .
\end{align*}
This shows that the Vainshtein mechanism operates in the region $r_*\ll r\ll r_V$. 
Furthermore, corrections to the GR expression for $\omega$ are suppressed by powers of $r_V$ in this region. 
On the other hand, the subleading correction to $\omega$ differs from GR in the region $r\leq r_*$, due to a different power law as compared to the GR case. 
Hence, the screening for $\omega$ is less effective in this region, meaning that corrections to the GR expression are not $r_V$ suppressed. One can check that the size of the value of $r_*$ is very small in physically relevant situations, i.e. $r_*\ll R$. 

\subsection{Cubic Galileon}
We now discuss the cubic Galileon theory, defined by (\ref{CG}). The static Vainshtein screening in this theory was studied in \cite{Babichev:2012re}, though the authors considered both a time dependence of the scalar field and a coupling of the scalar to the matter fields. The slow rotation in this theory has already been discussed in \cite{Chagoya:2014fza}, where they found that  the correction to the $(t\vp)$ equation coming from the Galileon term is highly suppressed. The scalar field equation (\ref{scalareq_alpha}) is quadratic in $\phi'$, and the solution reads
\begin{equation}
\label{phi_CG}
\phi' = \frac{2\alpha r_S r}{k_2 r_V^3}\left[1-\sqrt{1 + \frac{2GM r_V^3}{r_Sr^3}}\right]\; , 
\end{equation}
where we chose the solution that does not diverge at $r\to\infty$, and we defined $k_2 = \eta + 6 \alpha^2$ and $r_V^3 = 8\alpha\beta r_S/k_2^2$ (assuming $\alpha\beta>0$). 
The solution for the scalar field in the linear regime, $\phi'\sim \alpha r_S/(k_2 r^2)$, is similar to its counterpart in the case of k-essence~(\ref{philin}). The difference is that in the case of the cubic Galileon we included a canonical kinetic term; therefore, the limit $\alpha\to 0$ is well defined in this regime as well. In the linear regime, the equation for the frame-dragging function is modified in a similar way to the k-essence case~(\ref{omegaeqk}), and the conclusions of the previous section about a less effective screening for $\omega$ hold. Inside the Vainshtein radius, i.e. for $r\ll r_V$, we expand the solution (\ref{phi_CG}) and obtain
\begin{equation*}
\phi' = \frac{2\alpha r_S}{k_2 r_V^{3/2}}\sqrt{\frac{2GM}{ r_Sr}}\; .
\end{equation*}
In order to study the equation for the frame-dragging function inside the star, we assume the matter source has a constant density.
It is easy to check from the above expression that the fifth force is screened everywhere in the region $r\ll r_V$, 
unlike  in the k-essence case, where the fifth force becomes dominant for small radii inside the source (see Sec~.\ref{seckessence}). 
Substituting the expression for $\phi'$ into the $(t\vp)$ equation, we obtain the following expressions for the coefficients $K_1$ (outside and inside the source) and $K_2$:
\begin{align*}
	K_1^{\text{out}} & = \frac{4}{r}\left[1+ \mathcal{O}\left(\frac{r_S\sqrt{r}}{r_V^{3/2}}\right)\right]\; ,\\
	K_1^{\text{in}} & = \frac{4}{r}\left[1-\frac{3 r_S r^2}{8 R^3}\left(1+ \mathcal{O}\left(\frac{R^{3/2}}{ r_V^{3/2}}\right)\right)\right]\; ,\\
	K_2 &= -2 \left[1 + \frac{r_S r^2}{R^3}\left(1 + \mathcal{O}\left(\frac{R^{3/2}}{r_V^{3/2}}\right)\right)\right]\; .
\end{align*}
For a star of constant density, $\rho_0 = 3 r_S/R^3$, the leading corrections to the GR equation~(\ref{omegaweakGR}) are suppressed by powers of $r_V$. This means that the corrections to the GR solution for $\omega$ are also suppressed by powers of $r_V$, in a way analogous to the screening in spherical symmetry.

\subsection{Quadratic sector of Horndeski theory}
We now consider the quadratic sector of Horndeski theory, where $A_2=-A_1= f_X$, and $K=G_3=A_3=A_4=A_5=0$. We will treat the case $f_{XX}=0$ separately.
\subsubsection{Case \texorpdfstring{$f_{XX}\neq0$}{fXXneq0}}
For now let us assume $f_{XX}\neq 0$. Neglecting nonlinear terms in the $(tt)$ and $(rr)$ equations, the expressions for $\{\la,\nu\}$ are the same as Eq.~(\ref{kessence}). After substituting these expressions for the metric potentials, the scalar equation reads
\begin{equation}
\label{scalareqG4}
\alpha G M(r)+ 3 \alpha^2 r^2\phi'+\phi'^3 f_{XX}=0\; ,
\end{equation}
where $f_{XX}$ is evaluated at $X=-\frac{1}{2}\phi'^2$.
In the nonlinear regime outside the source, i.e. for $R\leq r \ll r_V$, the linear term in~(\ref{scalareqG4}) can be neglected. Then the scalar field is constant and verifies the equation,
\begin{equation*}
2\phi'^3 f_{XX}=-\alpha r_S\; ,
\end{equation*}
unless $f(X)\propto \sqrt{X}$, in which case the nonlinear term disappears in the scalar equation. For these particular theories, solving Eq.~(\ref{scalareqG4}) leads to $\p'\sim1/r^2$ everywhere outside the star. 
A similar case was studied in an application to black holes in \cite{Babichev:2017guv}.
In the general case, when $f(X)$ is not proportional $\sqrt{X}$, so that $f_{XX}\neq 0$, the derivative of the scalar field $\phi'$ must be constant.
This allows us to simplify the equation for $\omega$, since $\phi''=0$. 
Let us examine what happens for polynomial functions of the form
\begin{equation*}
f(X)=\frac{1}{2}+\kappa X^p\; ,
\end{equation*}
with $p>1$ so that $f_{XX}\neq 0$, and $\kappa$ is a constant coefficient. The spherically symmetric Vainshtein regime in such theories was discussed in~\cite{Kase:2013uja}. 
For large radii, one can neglect the nonlinear term in Eq.~(\ref{scalareqG4}), and the solution for $\phi'$ is the same as those for k-essence and the cubic Galileon discussed above.
One can then define a Vainshtein radius $r_V$ by equating the linear and nonlinear terms in Eq.~(\ref{scalareqG4}), and show that in the region $r \ll r_V$ the fifth force reads
\begin{equation}
\label{dphiquadratic}
\phi'\sim\frac{r_S}{6\alpha r_V^2} \left(\frac{2GM(r)}{r_S}\right)^{\frac{1}{2p-1}}\;  .
\end{equation}
The fifth force is constant outside the source, and one can easily check that it is screened for $r\ll r_V$.
Inside the source the situation is similar to the k-essence theories discussed above. 
Indeed, for a star of constant density 
the fifth force becomes larger than the Newtonian force near the center of the star when $p>2$ 
(for $p=2$, $|\phi'|$ grows linearly and the Vainshtein screening  is effective for all radii $r\ll r_V$).
For $p>2$, the fifth force becomes dominant for radii smaller than some $r_*$, which is much smaller than $R$.
A simple estimate, assuming that $r_V$ is of the order of the Neptune orbit, gives~$r_*\leq10$~m  (the case of k-essence is recovered for large $p$), while for more realistic Vainshtein radii, the value of $r_*$ is much smaller.
As in the k-essence theories, this small radius is not physically relevant.

Substituting~(\ref{dphiquadratic}) into the $(t\vp)$ equation, we obtain the following expressions for the coefficients $K_1$ (outside and inside the source) and $K_2$:
\begin{equation}
\label{G4_coeffs}
\begin{split}
K_1^{\text{out}} & =\frac{4}{r}\left[1 + \mathcal{O}\left(\frac{r_S r}{r_V^2}\right)\right]\; ,\\
K_1^{\text{in}} & = \frac{4}{r}\left[1 -\frac{3 r_S r^2}{8 R^3}\left(1 +\frac{R^2}{r_V^2}\cdot\mathcal{O}\left(\frac{R}{r}\right)^{\frac{2p-4}{2p-1}}\right) \right]\; ,\\
K_2 &= -2 \left[1 +\frac{ r_S r^2}{ R^3}\left(1 +\frac{R^2}{r_V^2}\cdot\mathcal{O}\left(\frac{R}{r}\right)^{\frac{2p-4}{2p-1}}\right) \right]\; .
\end{split}
\end{equation}
Using the results of section~\ref{sec:solutions}, one can see that for $p=2$ the situation is similar to the cubic Galileon case. The corrections to the GR expression for $\omega$ are suppressed by powers of $r_V$, and the screening operates in a way analogous to the spherically symmetric mechanism. For $p>2$, the situation is similar to the k-essence case, and the subleading terms in the solution for $\omega$ are not the same as in GR in the region $r\leq r_*$. However, as we discussed above, this region is not physically relevant.

\subsubsection{Case \texorpdfstring{$f_{XX}=0$}{fXX0}}
Let us now look at the case where the Lagrangian contains a derivative coupling to the Einstein tensor $\sim \phi^\mu\phi^\nu G_{\m\n}$, which corresponds to $f(X)=1/2+\kappa X$. 
The spherically symmetric Vainshtein mechanism in spherical symmetry in this theory was discussed in \cite{Koyama:2013paa}.
The particularity of this Lagrangian in application to the Vainshtein mechanism is that the leading nonlinear term in the scalar equation (\ref{scalareqG4}) vanishes. 
Therefore we have to keep nonlinear terms in the metric equations, as well as the subleading term for the scalar current, since the leading term vanishes. This modifies the expression for $\la$ (compared to (\ref{kessence})), and the metric potentials read
\begin{equation}
\label{john}
\begin{split}
\la &= \frac{2 G M(r)}{r}+2 \alpha r \p' + 2 \kappa \phi'^2\; ,\\
\nu' &=\frac{2G M(r)}{r^2}- 2\alpha\p'\; .
\end{split}
\end{equation}
Substituting these expressions into the scalar equation, we obtain
\begin{equation*}
\alpha G M(r) + 3\alpha^2 r^2 \phi'\left[1 +2\frac{\kappa\phi'}{\alpha r} + \frac{2}{3}\left(\frac{\kappa\phi'}{\alpha r}\right)^2\right]=0\; .
\end{equation*}
Defining the Vainshtein radius as $\kappa\phi'(r_V)\sim\alpha r_V$, which implies $r_V^3\sim\kappa r_S/\alpha^2$, both nonlinear terms are of the same order around $r\sim r_V$.
In the nonlinear regime $r\ll r_V$, the expressions~(\ref{john}) for the metric potentials imply that $\kappa r\phi'^2\ll GM$ in order for the static Vainshtein screening to work. In this case, one can show that the cubic term dominates in the scalar equation (otherwise we find $\kappa r\phi'^2\sim GM$, which modifies the GR expression for $\la$ in Eq.~(\ref{john})), and the fifth force reads
\begin{equation*}
\phi'\simeq-\frac{r_S}{\alpha r_V^2}\left(\frac{GM(r)}{2 r_S}\right)^{1/3}\; .
\end{equation*}
The above expression is similar to the one obtained for $p=2$ in the previous section. This means that the fifth force is screened for all radii $r\ll r_V$, inside and outside the matter source. After substituting this expression in the $(t\vp)$ metric equation, we obtain the following coefficients for the frame-dragging equation:
\begin{equation*}
\begin{split}
K_1^{\text{out}} & =\frac{4}{r}\left[1 + \mathcal{O}\left(\frac{r_S r}{r_V^2}\right)\right]\; ,\\
K_1^{\text{in}} & = \frac{4}{r}\left[1 -\frac{3 r_S r^2}{8 R^3}\left(1 +\mathcal{O}\left(\frac{R^2}{r_V^2}\right) + \mathcal{O}\left(\frac{r_S r^2}{R^2r_V}\right)\right) \right]\; ,\\
K_2 &= -2 \left[1 +\frac{ r_S r^2}{ R^3}\left(1 +\mathcal{O}\left(\frac{R}{r_V}\right)\right) \right]\; .
\end{split}
\end{equation*}
The subleading corrections depend on the value of $r$ inside the star. 
In any case, however, the corrections to the GR expression for $\omega$ are screened by a power of $r_V$, and the conclusions are the same as for $p=2$ in the previous section.
It is worth stressing again that in addition to these corrections due to modifications of gravity, there exist nonlinear GR terms. Both types of contributions can be seen as higher order corrections to linearized GR. We do not consider them here, though it is possible for these corrections to be larger than those coming from modified gravity.

\section{Conclusion}
\label{sec:conclusion}
We have studied the Vainshtein mechanism for slowly rotating stars in scalar tensor theories belonging to the DHOST Ia class. 
While the Vainshtein screening is usually studied for spherically symmetric objects, we have shown that in general slow rotation does not spoil  the Vainshtein screening. 
In the weak-field approximation the form of the leading term in the solution for the frame-dragging function coincides with the GR expression outside the star (up to an overall constant that can be reabsorbed in the definition of the angular momentum of the star). 
The angular momentum, being a constant of integration in the solution for the frame-dragging function, can be found if the mass distribution of the star is known.  
Inside the star, the Vainshtein screening can be broken for some theories, when the coefficient $K_2$ in the equation for the frame-dragging function~(\ref{tpeq}) receives leading order corrections.  
We also found that in most situations, when the Vainshtein screening operates in spherical symmetry, the leading corrections to the GR expression for $\omega$ in the weak-field approximation are also suppressed by powers of the Vainshtein radius.

An important qualification is in order. 
Although the corrections to $\omega$ may receive sizable modifications (inside the star), nevertheless the metric functions $\nu$ and $\lambda$ are not modified.
This means that if the theory exhibits the Vainshtein mechanism in spherical symmetry, slow rotation does not change the Vainshtein suppression of non-GR corrections to the ``static'' part of the metric $\nu$ and $\lambda$, independently of the behavior of the frame-dragging function $\omega$.

In our approach we applied the Hartle-Thorne formalism for slowly rotating stars to the scalar-tensor theories of the DHOST Ia class, Eq.~(\ref{actionDHOST}), (\ref{L}). 
We considered both a time-dependent and a time-independent scalar field, Eq.~(\ref{phians}). 
For rotating sources, the metric~(\ref{HTmetric}) contains the frame-dragging function $\omega$, which takes slow rotation into account. 
Our main purpose in the paper was to study the equation for $\omega$ and compare the results with the standard GR case. 
We found the general equation for the frame-dragging function in DHOST Ia theories, Eq.~(\ref{tpeq}), with coefficients of the equation given in (\ref{K1full}) and (\ref{K2full}).

For slowly rotating relativistic sources in a subclass of Horndeski theory~(\ref{Horndeski}), 
we calculated exact expressions for the coefficients $K_1$ and $K_2$. We have shown that in vacuum the GR equation for the frame-dragging function is fully recovered, see Sec.~\ref{sec:relativistic}. 
The latter result also applies to the quadratic beyond Horndeski theories, namely, for the theories described by the action~(\ref{BHtheory}), the vacuum equation for the frame-dragging function is the same as in GR. 
This result can be extended  general DHOST Ia theories, with the additional assumption that the kinetic term for the solution has the constant value $X=q^2/2$. 

In the rest of the paper, Sections~\ref{sec:solutions},\ref{sec:weakfield1}, and~\ref{sec:weakfield2}, we assumed that the weak-field approximation~(\ref{cond1}) is valid.
In Sec.~\ref{sec:solutions}, we showed that outside the star the solution for the frame-dragging function $\omega$ is the same as in GR at leading order. Inside the source, the screening can be broken, in this case $\kappa_2\neq-2$, see Eq.~(\ref{epsilon_coeffs}) . 
We also computed corrections to to solution for $\omega$ assuming that the coefficients of the frame-dragging equation acquire small modifications.
In Sec.~\ref{sec:weakfield1} we studied the equation for the frame-dragging function for various subclasses of the DHOST Ia class.
We found the expressions for the coefficients $K_1$ and $K_2$ of the equation for $\omega$ in this approximation, (\ref{K1exp}) and (\ref{K2exp}). 
Outside the Vainshtein radius, the coefficient $K_1$ receives a correction suppressed by $r_S/r$ (and by $q^2r^2$ in the time-dependent case), Eq.~(\ref{K1lin1}). 
To study the region inside the Vainshtein radius, we considered different classes of theories case by case. 
In most cases, when the Vainshtein screening works in spherical symmetry, the corrections to the GR expression for $\omega$ are screened by powers of $r_V$, in a way analogous to what happens in the nonrotating case.
However, we have found a particular theory for which the suppression is not as effective, in this case the leading correction is suppressed by $r_S/r$ instead.
We also studied a different class of theories for which the static metric potentials in the nonrotating case are exactly the same as in GR (possibly up to a redefinition of Newton's constant), while the screening for $\omega$ is broken inside the star. 

In the case of a static scalar field, see Sec.~\ref{sec:weakfield2}, the results are quite similar to the time-dependent case. 
In all the examples we considered, the Vainshtein mechanism works for the frame-dragging function $\omega$. 
Furthermore, the screening is more effective in regimes where the Vainshtein mechanism operates in spherical symmetry, meaning that the corrections to the GR expression are suppressed by powers of $r_V$. Meanwhile, outside the Vainshtein radius, the coefficients of the frame-dragging equation receive nonscreened corrections, see e.g. Eq.~(\ref{omegaeqk}) for k-essence. 
The screening still works for the frame-dragging function $\omega$, but it is less effective in this region.

Although the results of the paper show that the deviations from GR are always small (outside the source), it is interesting to see whether local gravity tests can provide additional constraints on scalar-tensor theories coming from the subleading modifications to the frame-dragging function. Probably the simplest way is to check constraints on PPN parameters (although it should be noted that precisely speaking the PPN analysis does not apply). The frame-dragging function $\omega$ can be written as (see e.g. section 4.4 of \cite{Will:2014kxa}),
\begin{equation*}
\omega_{\text{PPN}} = \left(1+\gamma +\frac14\alpha_1\right)\frac{J}{r^3}\; .
\end{equation*}
The deviations from GR are characterized by the combination $\gamma-1+\alpha_1/4$.
This is to be compared to our results on the frame-dragging function. 
Generically the deviation of $\omega$ from its GR value is of order $r_S/r$ for non-Vainshtein suppression, and much smaller for the Vainshtein suppressed cases. Therefore the combination of PPN parameters $\gamma-1+\alpha_1/4$ is not larger than $\mathcal{O}(r_S/r)$ in our case, which gives a deviation of order $10^{-8}$ at Earth's orbit. This value is well within the experimental constraints on both $\gamma$ and $\alpha_1$; therefore, we do not get any additional constraints on the parameters of the scalar-tensor theories from this estimation.
However, it would be worthwhile to look for a way to constrain particular classes of scalar-tensor theories by present or future observations using the results of this paper.}

\section*{Acknowledgments}
We would like to thank Gilles Esposito-Far\`ese for useful discussions. The work was supported by the CNRS/RFBR Cooperation program for 2018-2020 n. 1985 ``Modified gravity and black holes: consistent models and experimental signatures''.

\appendix
\section{List of coefficients}
\label{appendix:a}

We list here the coefficients of equations (\ref{ysol}), (\ref{zsol}), (\ref{scalarIa}), and (\ref{K1Ia}). Each time a function is written, it is evaluated on the time-dependent background. For instance,
\begin{equation*}
f\equiv f\left(q t,\frac{q^2}{2}\right)\; .
\end{equation*}
The time dependence of these coefficients comes from the $\phi$ dependence of the functions. We will implement the constraint $A_2= -A_1$, but in order  to keep expressions light, we will not substitute the expression for $A_4$ in DHOST Ia theories. One must keep in mind that the following constraint holds:
\begin{equation*}
\begin{split}
A_4 = & \frac{1}{8(f+q^2 A_1)^2} \left[12 f f_X^2+16q^2 A_1^3+(12 f + 32 q^2 f_X)A_1^2+(24 f f_X + 8 q^4 A_3 f_X + 16 q^2 f_X^2 - 12 q^2 f A_3)A_1 \right. \\
 &\left. + (4 q^2 f_X - 8 f - q^4 A_3) A_3 f \right]\, .
\end{split}
\end{equation*}
The terms involving $A_5$ were negligible in the field equations when assuming dimensionless quantities to be of $\mathcal{O}(1)$, so it does not appear in the following. 
\subsection{Coefficients of the metric equations}
With the definition
\begin{equation*}
C = f(2f + 2 q^2 A_1-q^4 A_4) + 2 q^2 f_X(q^2 f_X-f)\; ,
\end{equation*} the coefficients for equations (\ref{ysol}) and (\ref{zsol}) read: 
\begin{equation*}
\begin{split}
C\alpha_1  = & q^2 f\; ,\\
C\alpha_2  = & q^2\left(f-q^2f_X\right)\; ,\\
2C\beta_1 = & -2 f_\phi \left(f-2 q^2 f_X\right)-q^2\left[fG_X + q^2 f A_{3\phi} + \left(6f - 4 q^2 f_X\right)A_{1\phi}\right]\; ,\\
2C\beta_2 = &  q^2\left[\left(q^2 f_X - f\right)G_X - 2\left(f-2 q^2 A_1 - 3 q^2 f_X + q^4 A_4\right)A_{1\p} + \left(q^2 f_X - f\right)q^2 A_{3\phi}\right] \\
 &+ 2f_\p\left(f + 2 q^2 A_1 + q^2 f_X - q^4 A_4\right) \; ,\\
2C\gamma_1  = & 2 f\left( A_1 - q^2 A_{1X}\right) + q^2 f \left(3 A_3 + 2 A_4\right)  + 2 f_X \left(f - 2 q^2 f_X - 3 q^2 A_1\right) \; ,\\
2C\gamma_2 = &  A_1\left(3 q^4 A_4 - 4 f\right) + q^2\left(3 fA_3 + 2 f A_4 - 6 A_1^2\right) - f_X\left(2f + 6 q^2 A_1 + 3 q^4 A_3 + 2q^2 f_X\right) 
\\ &+ 2q^2A_{1X}\left(q^2 f_X - f\right)   \; ,\\
2C\delta_1 = & 2\left(A_1 + f_X\right)\left(f-2q^2f_X\right) + q^2 f\left(A_3 + 2 A_4\right) \; ,\\
2C\delta_2 = & q^2 f\left(A_3 + 2 A_4\right) - 2 A_1\left(f + 2 q^2 A_1-q^4 A_4\right)-f_X\left(2 f +2q^2 f_X+ 6 q^2 A_1 + q^4 A_3\right) \; ,\\
6C\eta_1 = & 2 f K + q^2\left[f\left(K_X-4 G_\phi\right) - 3 f_X\left(K-q^2 G_\phi\right)\right] \; ,\\
12C\eta_2 = & K\left(3 q^4 A_4 - 6 q^2 A_1 - 2 f\right) + q^2\left[2 f K_X - 2 f_X\left(2 K + q^2 K_X - 4 q^2 G_\phi\right) + G_\phi\left(6q^2 A_1-2f-3q^4 A_4\right)\right]\; .
\end{split}
\end{equation*}
\subsection{Coefficients of the scalar equation}
We now list the coefficients of the scalar equation (\ref{scalarIa}). We do not write $C_1$ or $\eta_3$ because the expressions are cumbersome, and we always neglect those terms in the nonlinear regime where the Vainshtein mechanism is operational. The other coefficients read
\begin{align*}
C_2& =  -6\{f+q^2 A_1\} \{A_1^3[36 q^4 f_\p - 24 q^6 f_{X\p}] \\
 & - 4 A_1^2 \big[3 q^6 f_\p\left(A_3-2A_{1X}\right) + f_X\left(q^6 G_{3X} + 6q^6 A_{1\p} + 4 q^4 f_\p + 4q^8A_{3\p} - 4 q^8 A_{1X\p}\right) \\
 &\; + q^8 f_{X\p} \left(4 A_{1X}-3A_3\right) - f\left(6 q^4 G_{3X} + 9 q^4 A_{1\p} + 12 q^6 A_{3\p} + 6 q^2 f_\p - 6 q^6 A_{1X\p} - 10 q^4 f_{X\p}\right)\big]\\
 &\; + q^2 A_1\big[ 4 q^4 A_3 \left(q^2 f_\p A_{1X} + 7 q^2f_X A_{1\p} + 5 f_X f_\p\right)  -3 q^6 f_\p A_3^2 + 4q^2 f_X\left(f_X[6q^2 A_{1\p} + 5 f_\p] - 2 A_{1X}[4 q^4 A_{1\p}+3q^2 f_\p]\right)   \\
 &\; + 8f^2\left(5 G_{3X} + 12 A_{1\p} + 10 q^2 A_{3\p} - 5 q^2 A_{1X\p} - 2 f_{X\p}\right) \\
 &\;-4 A_3 f\left( 2 q^4 G_{3X} + 15 q^4 A_{1\p} + 2 q^6 A_{3\p} + 10 q^2 f_\p + q^6 A_{1X\p} - 5 q^4 f_{X\p}\right)\\
 &\; + 4fA_{1X}\left(3 q^4 G_{3X} + 12 q^4 A_{1\p} + 4 q^6 A_{3\p} + 8 q^2 f_\p - 6 q^4 f_{X\p}\right) \\
  & - 4 f f_X\left(3 q^2 G_{3X} + 28 q^2 A_{1\p} + 8 q^4 A_{3\p} + 18 f_\p - 6 q^4 A_{1X\p}\right) ]\\
 &\; + f[A_3^2\left(5q^8 A_{1\phi}+ 2 q^6 f_\p\right)-4A_3(q^6 A_{1X}[2q^2 A_{1\p} + f_\p]-2q^4 f_X[5q^2A_{1\p} + 4 f_\p\big] + f\big[2q^4G_{3X}+16q^4 A_{1\p} + 2q^6 A_{3\p}\\
 &\; + 8 q^2 f_\p + q^6 A_{1X\p}-2q^4 f_{X\p}]) + 4 q^2 f_X\left(f_X[7q^2 A_{1\p} + 6 f_\p]-2 A_{1X}[6q^4 A_{1\p}+ 5 q^2 f_\p]\right)\\
& -8 f f_X(6f_\p + q^2[G_{3X} + 10A_{1\p} +q^2(2A_{3\p}-A_{1X\p})])+16 f^2\left(G_{3X} + 3A_{1\p}+q^2[2A_{3\p}-A_{1X\p}]\right)\\
 &\;+4q^2fA_{1X}\left(3q^2 G_{3X}+16q^2 A_{1\p}+4q^4A_{3\p}+ 6f_\p - 2q^2f_{X\p}\right)\big] \}  \; ,\\
C_3& =  24\left(f+q^2A_1\right)^2\left(f A_{1X}+A_1 f_X - f A_3\right)\left(2q^2 f_X + 3 q^4 A_3 - 4 f - 6 q^2 A_1 - 4 q^4 A_{1X}\right)\; ,\\
\Gamma_0 & =  -24 q^2\{f+q^2A_1\}\{f\left[q^2\left(2 f_X + q^2A_3\right)\left(2q^2 A_{1\phi}+f_\p\right)- f\left(2 f_\p + q^2\left(G_{3X} + 4 A_{1\phi}+ 2q^2 A_{3\p}-2f_{X\p}\right)\right)\right] \\& \;+A_1\left[ 2 q^4 f_{X\p}\left(3 f + 2q^2 A_1\right) - q^2 f_\p\left(4 f + 6 q^2 A_1 + q^4 A_3 + 2 q^2 f_X\right) - q^4 f G_{3X} - 2q^6 f A_{3\p}\right] \}     \; ,\\
\Gamma_1 & =  96 q^4\left(f + q^2 A_1\right)^2\left(f A_{1X}+A_1 f_X - f A_3\right)\; ,\\
\Gamma_2 & =  6 q^2\left[2 f A_1 + q^2\left(4 A_1^2-fA_3\right) + 2f_X\left(f + 2 q^2 A_1\right)\right] \left[f\left(4f+6q^2 A_1 + q^4 A_3\right) - 2 q^2 f_X\left(f + 2q^2 A_1\right)\right]  \; .\\
\end{align*}
\subsection{Coefficients of the \texorpdfstring{$(t\vp)$}{tvp} equation}
We now list the coefficients of Eq.~(\ref{K1Ia}), apart from $\beta_0,\kappa_0$, since we neglect these terms inside the Vainshtein radius. We define
\begin{equation*}
D=f\left(4f + 6 q^2 A_1 + q^4 A_3\right)- 2 q^2 f_X\left(f + 2 q^2 A_1\right)\; .
\end{equation*}
 The remaining coefficients read
\begin{align*}
D^2\alpha_0 & = -2 q^4 \left(f + q^2 A_1\right)\left(2 f A_1 + q^2 f A_{1X} + f_X\left[2f + q^2 A_1\right]\right)\; ,\\
D^2\zeta_0 & = -2 q^2 \left(f + q^2 A_1\right)^2\left(f-q^2 f_X\right)\; ,\\
4D^2\gamma_0 & = \{f+q^2A_1\}\{f\left[4q^2 A_{1X}\left(2f + 4q^2 A_1 + 2 q^4 A_{1X}-3q^4 A_3\right)+3q^6 A_3^2 -12q^2\left(A_1^2 + f A_3\right)-8fA_1 -24q^4A_1A_3\right]\\ 
& \qquad + 8 f_X\left[3 q^4 A_1^2 + q^6 A_1 A_{1X} - f^2\right] + 4 q^2f_X^2\left[f+2q^2 A_1\right]\} \; ,\\
2D^2\delta_0 & = \left(f+q^2 A_1\right)\left(2 A_1 + q^2 A_3 + 2 f_X\right)\left(f\left[4f + 6 q^2 A_1 + 3 q^4 A_3 - 2 q^4 A_{1X}\right]- 2q^2f_X\left[f + 3 q^2 A_1\right]\right) \; ,\\
4D\sigma_0 & = \left(f+q^2 A_1\right)\left(2 A_1 + q^2 A_3 + 2 f_X\right)  \; .\\
\end{align*}
\subsection{Other coefficients}
We define
\begin{equation*}
B = 16\left(f+ q^2 A_1\right)\left(A_1 f_X + f A_{1X} - f A_3\right)\left[4f + 2 q^2\left(3 A_1 - f_X\right) + q^4\left(4 A_{1X} - 3 A_3\right)\right]\; .
\end{equation*}
Then, the coefficients of Eq.~(\ref{K1x1}) read
\begin{align*}
B\iota_0 =& -4 q^2\left(f + 2 q^2 A_1\right) \left[A_1^2-2f A_3 + f_X \left(4 A_1 + f_X\right)\right] - q^6 A_3\left(4 A_1^2 + 3 fA_3\right) - 8 f q^6 A_{1X}^2 \\
& - 4 q^2 A_{1X}\left[2 q^4 A_1 f_X + f\left(2f + 4 q^2 A_1 - 3 q^4 A_3\right)\right] \; ,\\
B\iota_1 =& q^2\left(2 A_1 + 2 f_X + q^2 A_3\right)\left[2 fA_1 + 2 f_X\left(f + 3q^2 A_1\right) + q^2\left(4A_1^2 - 3fA_3 + 2 f A_{1X}\right)\right] \; , \\
2B\iota_2 =& q^2\left(2 A_1 + 2 f_X + q^2 A_3\right)\left[2 fA_1 + 2 f_X\left(f + 2q^2 A_1\right) + q^2\left(4A_1^2 - fA_3\right)\right] \; .\\
\end{align*}
The coefficient of Eqs.~(\ref{ycex}) and (\ref{zcex}) reads
\begin{align*}
	\iota_3 &= \frac{4 A_{1\phi}\left[6 + 4 q^2 A_1\left(7 + 10 q^2 A_1 + 6 q^4 A_1^2\right)- q^4 A_{1X}\left(2 + 10 q^2 A_1 + 4 q^4 A_1^2 + q^4 A_{1X} \right)\right]}{2\left(2 + 6 q^2 A_1 + q^4 A_{1X}\right)^2\left[3 A_{1\phi}\left(2 + 2q^2 A_1 -  q^4 A_{1X}\right)+2\left(1 + 2 q^2 A_1\right)\left( G_{3X} +  q^2 A_{1\phi X}\right)\right]}\\
	& + \frac{ \left(1 + 2 q^2 A_1\right)\left[G_{3X}\left(8 + 30 q^2 A_1 + 24 q^4 A_1^2 + q^4 A_{1X}\right) + 2 q^2 A_{1\phi X}\left(4 + 14 q^2 A_1 + 8 q^4 A_1^2 + q^4 A_{1X}\right)\right]}{2\left(2 + 6 q^2 A_1 + q^4 A_{1X}\right)^2\left[3 A_{1\phi}\left(2 + 2q^2 A_1 -  q^4 A_{1X}\right)+2\left(1 + 2 q^2 A_1\right)\left( G_{3X} +  q^2 A_{1\phi X}\right)\right]}\; .
\end{align*}
\section{Relation between Schwarzschild and Newtonian potentials}
\label{appendix:b}
We briefly remind the reader of the way to switch between the functions $\{\la,\nu\}$ used throughout this work to the Newtonian potentials $\{\Phi,\Psi\}$ often encountered in the literature. The two line elements we want to relate in the weak-field limit are (we set $\omega\simeq 0$)
\begin{align*}
\td s^2& = -\left(1+\nu(r) \right)\td t^2 + \left(1+ \la(r)\right)\td r^2 + r^2 \td \Omega^2\; ,\\
\td s^2& = -\left(1+2\Phi(\bar{r}) \right)\td t^2 + \left(1-2 \Psi(\bar{r})\right)\left[\td \bar{r}^2 + \bar{r}^2 \td \Omega^2\right]\; .
\end{align*}
Then, in the Newtonian limit $\{\Psi,\Phi\}\ll 1$ , we obtain
\begin{gather*}
r\simeq\bar{r}\left(1-\Psi\right)\simeq \bar{r}\; ,\\
\nu = 2\Phi\; ,\\
\la = 2 r \Psi'\; .
\end{gather*}

\end{document}